\documentstyle[aps,epsf,eqsecnum,tighten,preprint]{revtex}
\begin{document}
\title{Universal relaxational dynamics near two-dimensional
quantum-critical points}
\author{Subir Sachdev}
\address{Department of Physics, Yale University\\
P.O. Box 208120,
New Haven, CT 06520-8120, USA}
\date{Oct 31, 1998}
\preprint{cond-mat/9810399}
\maketitle
\begin{abstract}
We describe the nonzero temperature ($T$), low frequency ($\omega$)
dynamics of the order parameter near quantum critical points in
two spatial dimensions ($d$), with a special focus on the regime
$\hbar \omega \ll k_B T$. For the case of a `relativistic', ${\rm
O}(n)$-symmetric, bosonic quantum field theory we show that, for small
$\epsilon=3-d$, the dynamics is described by an effective
classical model of {\em waves} with a quartic interaction.
We provide analytical and numerical analyses of the
classical wave model directly in $d=2$.
We describe the crossover from the finite
frequency, ``amplitude fluctuation'', gapped quasiparticle mode in the quantum
paramagnet (or Mott insulator),
to the zero frequency ``phase'' ($n \geq 2$) or ``domain wall'' ($n=1$)
relaxation mode
near the ordered state.
For static properties, we
show how a surprising, duality-like transformation allows
an exact treatment of the strong-coupling limit for all $n$.
For $n=2$, we compute the universal $T$ dependence of the
superfluid density below the Kosterlitz-Thouless temperature,
and discuss implications for the
high temperature superconductors. For $n=3$, our computations of
the dynamic structure factor relate to neutron scattering
experiments on ${\rm La}_{1.85} {\rm Sr}_{0.15} {\rm Cu O}_4$,
and to light scattering experiments on double layer quantum Hall
systems.
We expect that closely related effective classical wave models
will apply also to other quantum critical points in $d=2$.
Although computations in appendices do rely upon technical results on the
$\epsilon$-expansion of quantum critical points
obtained in earlier papers, the physical discussion in the body of the paper
is self-contained, and can be read without consulting
these earlier works.
\end{abstract}
\pacs{PACS numbers:}
\newpage
\section{Introduction}
\label{intro}

A number of recent experiments have probed the long-wavelength,
low frequency, nonzero temperature ($T$) dynamics of the order
parameter associated with a $T=0$ quantum critical point in
a two spatial dimensions ($d$). These experiments include:
\newline
({\em i\/}) Neutron scattering measurements have mapped out the
$T$, wavevector, and frequency dependence of the dynamic spin
structure factor in ${\rm La}_{2-x} {\rm Sr}_x {\rm Cu O}_4$ for
$x \approx 0.15$~\cite{aeppliscience}. The measurements over an order
of magnitude in $T$, and over three orders of magnitude in the static
susceptibility, are consistent with the presence of a nearby
quantum critical point to an insulating ordered state with
incommensurate spin and charge order (`stripes'~\cite{tran}).
\newline
({\em ii\/}) Double layer quantum Hall systems at filling factor $\nu = 2$
exhibit ground states with different types of magnetic order~\cite{DSZ}.
Recent light scattering experiments~\cite{pelle} have probed the fluctuation of
the magnetic order parameter in the vicinity of the quantum
transitions between the states.
\newline
({\em iii\/}) Microwave measurements~\cite{hardy} of the magnetic penetration depth
of high temperature superconductors with a number of different
$T_c$'s show that the superfluid stiffness satisfies the scaling
relation
\begin{equation}
\frac{\rho_s (T)}{\rho_s (0)} = \Psi_{\rho} \left( \frac{T}{T_c} \right),
\label{r1}
\end{equation}
where $\Psi_{\rho}$ is an apparently universal function. This is
precisely the behavior expected in the vicinity of a quantum
critical point where $T_c \rightarrow 0$~\cite{eps}, such as a
superfluid-insulator transition.
In this paper, we shall provide an explicit computation of the
universal scaling function $\Psi_{\rho}$ in a model system. We
believe our approach and strategy can be generalized to
models which include some of the additional physics
contained in recent discussions~\cite{ek,franz,dorsey}
of the $T$
dependence of the superfluid density in the high temperature
superconductors.

Motivated by these disparate experimental systems, this paper will
present an analysis of the long-wavelength, nonzero temperature
order-parameter
dynamics in the vicinity the simplest, interacting quantum critical point
in $d=2$: that of a `relativistic', $n$-component, bosonic field
$\phi_{\alpha}$, $\alpha= 1 \ldots n$. However, our ideas and
approach are expected to be far more general, as we shall
discuss further in Section~\ref{expts}. The ${\rm O}(n)$-symmetric
quantum partition function
for the field $\phi_{\alpha}$ is given by (in units with $\hbar = k_B = 1$
which we use throughout)
\begin{eqnarray}
&& {\cal Z}_Q = \int {\cal D} \phi_{\alpha} (x, \tau) \exp \left( - \int d^d x
\int_0^{1/T} d \tau \, {\cal L}_Q
\right) \nonumber \\
&& {\cal L}_Q =  \frac{1}{2} \left[
 \frac{1}{c^2} (\partial_{\tau} \phi_{\alpha})^2 + (\nabla_x \phi_{\alpha})^2 +
(r_c + r) \phi_{\alpha}^2 \right]
+ \frac{u}{4!} \left( \phi_{\alpha}^2 \right)^2 .
\label{calsq}
\end{eqnarray}
Here $x$ is the $d$-dimensional spatial co-ordinate, $\tau$ is
imaginary time, $c$ is a velocity, and $r_c$, $r$ and $u$
are coupling constants.
The co-efficient of the $\phi_{\alpha}^2$ term (the `mass' term)
has been written as
$r + r_c$ for convenience; we will choose the value of $r_c$ so
that the quantum critical point is precisely at $r=0$. So the $T=0$
ground state has spontaneous `magnetic' order for $r<0$ with
$\langle \phi_{\alpha} \rangle \neq 0$, and is a quantum
paramagnet with complete ${\rm O}(n)$ symmetry preserved for
$r>0$.
The quartic non-linearity proportional to $u$ is relevant about
the Gaussian fixed point ($u=0$) for $d<3$, and is responsible for
producing a non-trivial quantum critical theory with interacting
excitations. Higher-order non-linearities are irrelevant about
this quantum critical point.

In addition to being an important and instructive toy model of an
interacting quantum critical point in $d=2$,
the field theory (\ref{calsq}) also has direct applications to
experimental systems. We will briefly note these now, and discuss them
further in Section~\ref{expts}. For $n=2$, ${\cal Z}_Q$
describes the transition between superfluid and Mott-insulating
states of an interacting boson model: $\phi_1 + i \phi_2$ is
the superfluid order parameter and the quantum paramagnet is a
Mott insulator. For $n=3$, $\phi_{\alpha}$ plays the role of a
magnetic order parameter measuring the amplitude of the
incommensurate, collinear spin density wave in the experiments of
Ref.~\onlinecite{aeppliscience}. The $n=3$ case also describes the
quantum Hall experiments of Ref.~\onlinecite{pelle}, where $\phi_{\alpha}$
now measures the difference in the magnetization of the two
layers.

\begin{figure}
\epsfxsize=13in
\centerline{\epsffile{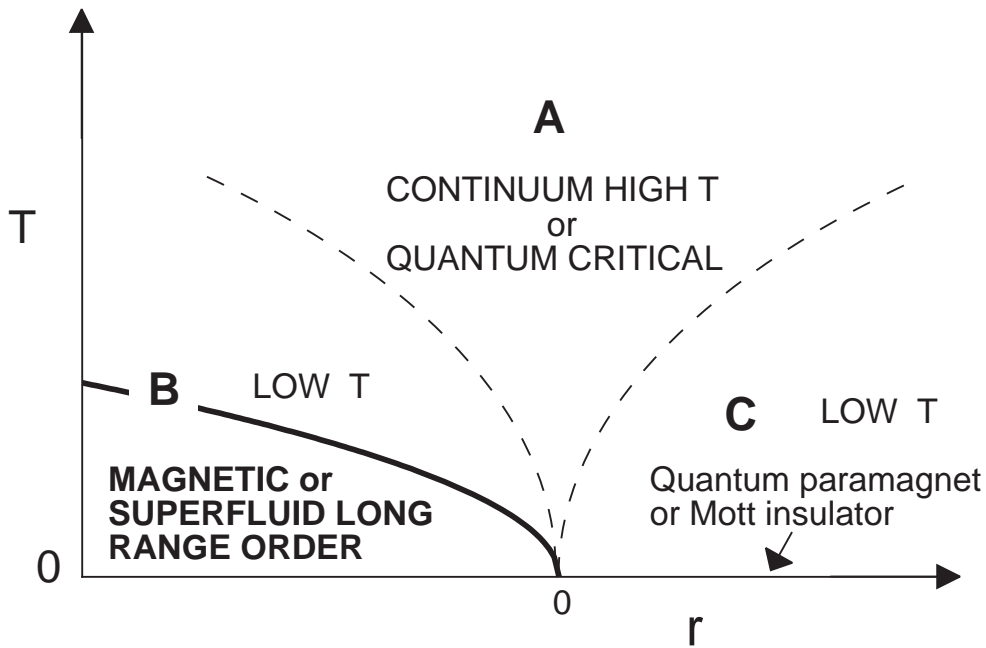}}
\caption{Phase diagram of the theory ${\cal Z}_Q$ for $d=2$,
$n=1,2$ as a function of the temperature $T$ and tuning parameter
$r$. The quantum critical point is at $T=0$, $r=0$.
The most important crossovers are represented
by the dashed lines, and these occur at $T \sim |r|^{z
\nu}$, where the dynamic exponent $z=1$, and $\nu$ is the
correlation length exponent of the $d+1$-dimensional classical
ferromagnet; these crossovers divide the phase diagrams into
regions A, B, and C. Region A is the high temperature of the
continuum theory ${\cal Z}_Q$, with the $T \rightarrow \infty$
limit taken {\em after} the short distance cutoff has been sent to
zero to obtain the continuum limit; its properties are described
by placing the $r=0$ scale-invariant critical theory at
non-zero temperature. There are two low $T$ regions, B, C, on
either side of $r=0$. The ground state for $r>0$ is a quantum
paramagnet (or a Mott insulator, depending upon the physical
system) with an energy gap; the dynamics in low $T$ region C is
described by a model of a dilute gas of thermally excited {\em
quasi-classical particles}, and this shall not be discussed in
this paper. The ground state for $r<0$ has long range order with
$\langle \phi_{\alpha} \rangle \neq 0$ and the low $T$ properties above
it are described by a model of {\em quasi-classical waves} for $n \geq 2$
(for $n=1$ a separate model of quasi-classical particles applies).
There is a line of finite
temperature phase transitions, $T=T_c (r)$, within region B at which the
long-range order disappears; this is denoted by the full line.
This $T>0$ transition is of the Kosterlitz-Thouless type for
$n=2$, and in the universality class of the two-dimensional
classical Ising model for $n=1$. The phase diagram for $n\geq 3$
differs only in that there is no line of $T>0$ phase transitions
in region B {\em i.e.} $T_c (r) = 0$, and long-range order is
present only for $T=0$, $r<0$. The present paper uses the $\epsilon=3-d$
expansion to develop a theory
for the low frequency ($\omega < T$), long distance dynamics in
region A
directly in $d=2$ for all $n$, using a model of {\em quasi-classical
waves} (our model also contains the initial crossovers as $T$ is
lowered into regions B or C).
In contrast, transport of the conserved ${\rm O}(n \geq 2)$ charge in region A
was discussed in II for small $\epsilon$; it was dominated by excitations with
energy $\varepsilon_k \sim
T$, and described by particles obeying a quantum transport
equation.
}
\label{fig1}
\end{figure}

An important tool in the analysis of ${\cal Z}_Q$ is the
$\epsilon$ expansion, where
\begin{equation}
\epsilon = 3-d.
\label{r2}
\end{equation}
The structure of the $\epsilon$ expansion for the
$T>0$ properties of ${\cal Z}_Q$ has already been extensively
discussed in two previous papers, hereafter referred
to as I~\cite{eps} and II~\cite{ds}.
We will now summarize the main results of these papers, and then
turn to a description of the specific purpose of this
paper. Although the present paper builds upon on these earlier
works, an attempt has been made to make all of the physical discussion
in the body of the paper self-contained;
earlier technical results are used in the appendices.
 Some physical results from I
are summarized in the caption of Fig~\ref{fig1}, which shall form the basis
of our subsequent discussion.

In I,
the properties of the phase diagram in Fig~\ref{fig1} were
analyzed in an expansion in $\epsilon$. In particular, detailed
$\epsilon$ expansion results were obtained for the dynamic
susceptibility
\begin{equation}
\chi( k ,  \omega_n ) \equiv \frac{1}{n} \int_0^{1/T} d \tau \int d^d x \,
\sum_{\alpha=1}^{n}
\langle \phi_{\alpha} (x, \tau) \phi_{\alpha} (0,0) \rangle e^{-i(kx - \omega_n
\tau)},
\label{defchi}
\end{equation}
where
$k$ is the wavevector, $\omega_n$ the imaginary
frequency; throughout we will use the symbol $\omega_n$ to refer to
imaginary frequencies, while the use of
$\omega$ will imply the expression has been analytically continued to real
frequencies. It was found in I that for the {\em static\/}
susceptibility
\begin{equation}
\chi (k) \equiv \chi(k, \omega_n = 0),
\label{defstat}
\end{equation}
an expansion in powers of $\sqrt{\epsilon}$ held over all regions
of the phase diagram of Fig~\ref{fig1}, apart from a small window
in the immediate vicinity of the line of finite temperature phase
transitions (in this window, the problem reduces to one in classical critical
phenomena, and this shall not be of interest to us here).
So in a sense, the static theory was weakly coupled
for small $\epsilon$, and this allowed for a satisfactory
theoretical treatment of the crossovers in $\chi (k)$.
A completely different situation held for the dynamic properties,
and in particular for the spectral density
$\mbox{Im} \chi (k, \omega)$: the $\epsilon$
expansion was found to fail badly, and led to unphysical results
for small $k$ and $\omega$. In particular, in region A, this
failure in the computation of the dynamic properties appeared for
wavevectors smaller than $c k \sim \sqrt{\epsilon} T$ and
frequencies smaller than $\omega \sim \sqrt{\epsilon} T$.
In the $\epsilon$ expansion, the low frequency spectral density is
given by an integral over the phase space for the decay of
excitations into multiple excitations at lower energy; however it
does not self-consistently include damping in these final
states, and this leads to unphysical results. In other words, determination of the
values of $\mbox{Im} \chi (k,
\omega)$, for small $k$ and $\omega$ requires solution of a strong
coupling problem, {\it even for small $\epsilon$},
and is dominated by the relaxation of excitations with energies of
order or smaller than $\sqrt{\epsilon} T$.
 Similar results
hold also for the expansion in $1/n$~\cite{SY,CSY}. It is this
strong coupling problem which will be addressed in this paper.

Although, as just discussed, the results of I for static
properties were adequately computed at low orders in the $\epsilon$
expansion, even they had significant qualitative weaknesses when
extrapolated to the physically interesting case of $\epsilon=1$,
$d=2$, apart from also not being quantitatively very accurate.
In particular, we know that the line of non-zero
temperature phase transitions in Fig~\ref{fig1} is not present for
$n \geq 3$ {\em i.e.} $T_c (r) = 0$ for these cases. In contrast,
leading order $\epsilon$ expansion results of I have a $T_c (r) > 0$
for all $n$. Furthermore, for $n=2$, there should be a jump in the
value of the superfluid density at $T_c$ in $d=2$, and clearly, this also
does not appear at any order in the $\epsilon$ expansion.
We shall address all of these problems in this paper, along with
the dynamical problem indicated above. We will do this by an exact treatment of
certain thermal fluctuations {\em directly in $d=2$}, while the
remaining quantum and thermal effects (for which the case $d=2$
plays no special role) are treated by a low order $\epsilon$
expansion. We shall claim that this hybrid approach leads to a
more
quantitatively accurate determination of both static and dynamic
properties in the high temperature (or quantum critical) region A
of Fig~\ref{fig1}. Our approach will lead to a computation
of the scaling function $\Psi_{\rho}$, in (\ref{r1}), containing
the universal jump in the superfluid density at $T_c$.

Before turning to a discussion of our strategy in solving the
strong-coupling dynamical problem, let us also review the results of II.
This paper examined transport of the conserved charge associated
with the continuous ${\rm O} (n)$ symmetry of ${\cal Z}_Q$, for $n \geq
2$ and $\epsilon$ small.
The $T$ and $r$ dependence of the conductivity was examined
using a perturbative expansion in $\epsilon$,
especially in region A. It was found that for small $\epsilon$,
the most important
current carrying states were bosonic particle
excitations with energy $\varepsilon_k \sim
T$, and momentum $k \sim T/c$ (contrast this with the typical energy
of order $\sqrt{\epsilon} T$ which dominates relaxation of the order
parameter, as discussed above, and in I).
The damping and scattering of
the current carrying states with $\varepsilon_k \sim T$
was adequately computed by the $\epsilon$ expansion
of I, as they were out of the region of $k, \omega$ space where
the weak-coupling expansion broke down. Also, because $\varepsilon_k$
was not much smaller than $T$, the occupation number of these
bosonic modes could not be approximated by the classical
equipartition value, $T/\varepsilon_k$, but required the full
function $1/(e^{\varepsilon_k/T} - 1)$ for quantized Bose particles.
The transport of charge by these particles was analyzed by the
solution of a quantum Boltzmann
equation in II. All the analysis of II was systematic in powers
of $\epsilon$, and included only the leading non-trivial terms.
The present paper will develop a new strong-coupling approach
in $d=2$,
but it will be applied only to the low frequency order
parameter dynamics; the transport properties of the new
approach will be examined in a future publication.

We are now ready to outline the strategy of this paper. We will
begin in Section~\ref{statics} by recalling the approach of I for
the computation of static properties in $\epsilon$ expansion.
We shall show that a straightforward modification of this approach
allows an exact treatment of the most singular thermal
fluctuations in directly in $d=2$, allowing us to obtain results
which have all the correct qualitative features for all values of
$n$, and are also believed to be quantitatively accurate.
The low frequency dynamic properties will then be considered in
Section~\ref{dynamics}.

\subsection{Statics}
\label{statics}

The main idea of I was to analyze ${\cal Z}_Q$ in two steps.
In the first, all modes $\phi_{\alpha} (k, \omega_n)$
with a non-zero frequency, $\omega_n \neq  0$ were integrated out
in a naive $\epsilon$ expansion.
This produced an effective action for the $\phi_{\alpha} (k, \omega_n = 0)$
modes, which was subsequently analyzed by more sophisticated
techniques. Our first step here will be identical to that in I; so
we define
\begin{equation}
\Phi_{\alpha} (x) \equiv T \int_0^{1/T} d \tau \, \phi_{\alpha} (x,
\tau).
\end{equation}
After integrating out the modes with a non-zero $\omega_n$, the
effective action for $\Phi_{\alpha} (x)$ takes the form
\begin{eqnarray}
&& {\cal Z} = \int {\cal D} \Phi_{\alpha} (x) \exp \left( - \frac{1}{T}
\int d^d x \, {\cal L}
\right) \nonumber \\
&& {\cal L} =  \frac{1}{2} \left[ (\nabla_x \Phi_{\alpha})^2 +
\widetilde{R} \, \Phi_{\alpha}^2 \right]
+ \frac{U}{4!} \left( \Phi_{\alpha}^2 \right)^2 .
\label{cal}
\end{eqnarray}
The values of the coupling constants $\widetilde{R}$ and $U$ will
be discussed shortly. We are treating the consequences of the
non-zero $\omega_n$ modes at the one loop level, and at this order
the co-efficient of the spatial gradient term does not get
renormalized. This approximation means that effects associated
with wavefunction renormalization and the quantum critical exponent $\eta$
have been neglected: this is quite reasonable, as $\eta$ takes
rather small values at the $2+1$ dimensional quantum critical
point. Also, these two loop effects were considered at length in
I, and were found to be quite unimportant.

We note in passing that ${\cal Z}$ in (\ref{cal})
is designed to apply
as a model of ${\cal Z}_Q$ in (\ref{calsq}) in region A
of Fig~\ref{fig1}. It also contains the initial crossovers into
regions B and C, but there are some subtleties as $T \rightarrow 0$
in regions B and C. As we shall see in Appendix~\ref{ru}, the limits
$T \rightarrow 0$ and $\epsilon \rightarrow 0$ do not commute: the
leading physics for very small $T$ can be properly captured, but there are some
subleading effects which are not accounted for in an approach
based on the $\epsilon$ expansion. These caveats also
apply to the dynamics to be discussed in Section~\ref{dynamics}.

\begin{figure}
\epsfxsize=4in
\centerline{\epsffile{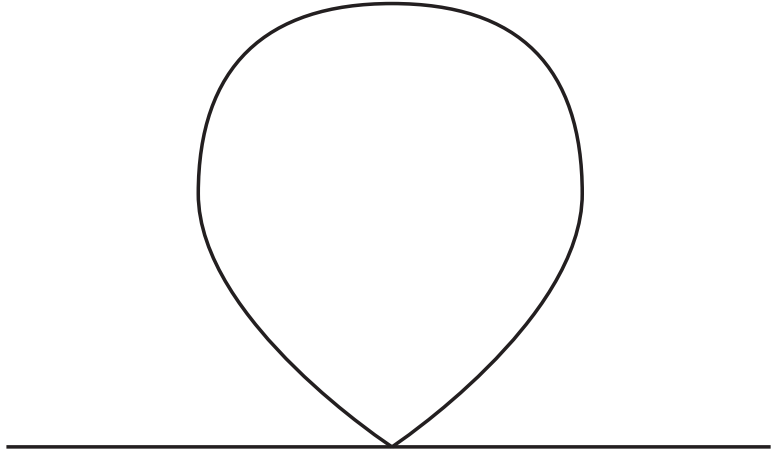}}
\caption{`Tadpole' graph containing the only ultraviolet
singularity of ${\cal Z}$}
\label{fig2}
\end{figure}

For our remaining discussion, it is crucial to understand the
properties of ${\cal Z}$ as a continuum, classical field theory in
its own right. Our strategy here will be obtain the universal
properties of this continuum theory {\em directly in $d=2$.}
Actually (\ref{cal}) does not define the theory ${\cal Z}$
completely, as some short distance regularization is needed to
remove the ultraviolet divergences. A priori, it might seem that
there is no arbitrariness in choosing the short distance
regularization, as it is uniquely provided by the underlying quantum theory
${\cal Z}_Q$. However, as will become clear now, it is actually
possible to choose a `virtual' short distance regularization of ${\cal Z}$
at our convenience, provided we properly match certain
renormalized couplings with those obtained from the true quantum
regularization due to ${\cal Z}_Q$; we will work with a `virtual'
lattice regularization of ${\cal Z}$ here.
So, what are the short distance singularities of ${\cal Z}$~? From
standard field theoretic computations~\cite{ramond}, it is known
that for $d<3$, the model ${\cal Z}$ has only {\em one}
ultraviolet divergence coming from the `tadpole' graph shown in
Fig~\ref{fig2} (there are some additional divergences, associated with composite
operators,
which appear when two or more field operators approach each other in space: we
will not be concerned with these here---however, these are important in a
consideration of transport properties, and will be discussed in a future
paper).
So all short scale dependence can be removed
simply by defining a renormalized coupling $R$ related to the bare
coupling $\widetilde{R}$ by
\begin{equation}
\widetilde{R} = R - T U \left(\frac{n+2}{6} \right) \int^{1/a}
\frac{d^d k}{(2 \pi)^d} \frac{1}{k^2 + R}.
\label{defR}
\end{equation}
Here the expression $1/(k^2 + R)$ should be read as a schematic
for the low momentum behavior of the propagator. At higher momenta
of order $1/a$ (for lattice regularization, we will choose $a$ to
be the lattice spacing), the propagator can have a rather
different momentum dependence and this has to be accounted for
in computing the integral in (\ref{defR}). Also notice that we
have performed the subtraction with a propagator carrying the
renormalized `mass' $R$. For $d>2$, this is not crucial and we can
equally well define the subtraction with a massless propagator
$1/k^2$: this procedure was followed in I, and has the
advantage of leading to a simple linear relation between $\widetilde{R}$
and $R$. Here, we are interested in working directly in $d=2$, and
 then such a massless subtraction would lead to an infrared
 divergence. So we are forced to perform the subtraction as in
 (\ref{defR}). Indeed in $d=2$,
(\ref{defR}) evaluates to
\begin{equation}
\widetilde{R} = R - T U \left(\frac{n+2}{24 \pi} \right)
\ln(C/Ra^2),~~~~~~~~~~~~~d=2
\label{defrt}
\end{equation}
where $C$ is a regularization dependent, non-universal constant.
This clearly shows that it is not possible to set $R=0$ in the
subtraction term. We also note an important property of
(\ref{defR}, \ref{defrt}) which is special to $d=2$.
Clearly, we have assumed above that $R>0$. However (\ref{defrt})
shows that the bare mass $\widetilde{R}$ ranges from $-\infty$
to $+\infty$ as $R$ increases from 0 to $\infty$. So it is no
restriction to consider only
positive values of $R$, as that allows us to scan the bare mass in ${\cal Z}$
over all possible negative and positive values (this is not true for $d>2$ as the
reader can easily check from (\ref{defR}):
then we do need values of $R<0$, while defining the renormalization with a
massless propagator, to access all values of $\widetilde{R}$). In particular, as we will
show, we will be able to access both the magnetically ordered and disordered phases
of ${\cal Z}$ for $R>0$ in $d=2$.
We can also interpret (\ref{defrt}) in a renormalization group
sense as defining the scale-dependent effective mass $\widetilde{R}$
at a length scale $a$, in a theory with a fixed, positive $R$:
so even in a theory with $R>0$, it is possible to have a
significant window of length scale, where the scale-dependent mass
$\widetilde{R}$ is less than zero.
We will see in Sections~\ref{dyn} and~\ref{expts} that this
interpretation is very helpful in understanding the origin of
`pseudo-gap' physics in the quantum critical region.

After short distance dependencies have been removed by the simple
renormalization in (\ref{defrt}), all correlators of ${\cal Z}$ are expected to
be finite in the limit $a \rightarrow 0$, and are universal
functions of the renormalized couplings $R$ and $U$.
Actually, instead of working with $R$ and $U$, we shall find it
more convenient to use $R$, and the dimensionless {\em Ginzburg parameter},
${\cal G}$, defined by
\begin{equation}
{\cal G} \equiv \frac{ T U }{R^{(4-d)/2}},
\label{deflam}
\end{equation}
as our two independent couplings; the ratio ${\cal G}$ gives an
estimate of the strength of the non-linear fluctuations about the
mean-field plus Gaussian fluctuation treatment of ${\cal Z}$.
So, provided
we express everything in terms of $R$ and ${\cal G}$ (and not $\widetilde{R}$
and $U$), the properties of ${\cal Z}$ are regularization
independent and universal functions of $R$ and ${\cal G}$.

At this point, the correct approach towards computing the static
properties of the underlying quantum problem ${\cal Z}_Q$ should
be quite clear.
\newline
({\em i\/}) First, we compute the values of the effective
couplings $R$ and ${\cal G}$, defined in (\ref{cal}), (\ref{defR})
and (\ref{deflam}),
by integrating out the nonzero $\omega_n$ modes in (\ref{calsq}).
This is carried out by an extension of the approach developed
in I, and our results are presented in
Appendix~\ref{ru}.
In the most interesting high $T$ region A of Fig~\ref{fig1}, these couplings
take the following values to leading order in $\epsilon$
\begin{eqnarray}
R &=& \epsilon \left( \frac{n+2}{n+8} \right) \frac{2 \pi^2
(T/c)^2}{3} \nonumber \\
{\cal G} &=& \sqrt{\epsilon} \frac{48 \pi \sqrt{3}}{\sqrt{2 (n+2)(n+8)}}.
\label{rest}
\end{eqnarray}
As one moves out of the region A, these couplings become smooth,
monotonic, and universal functions of $r/T^{1/z\nu}$. In
particular, ${\cal G}$ obeys the simple scaling form
\begin{equation}
{\cal G} = \Psi_{\cal G} \left( C_1 \frac{r}{T^{1/z\nu}} \right),
\label{phig}
\end{equation}
where $\Psi_{\cal G}$ is a universal function, and $C_1$ is a
non-universal scale which can be eliminated if the argument
of $\Psi_{\cal G}$ is related to the actual $T=0$ energy gap or
spin stiffness (as discussed in I and Appendix~\ref{ru}).
As one moves in Fig~\ref{fig1} from Region C to A to B, the
argument of $\Psi_{\cal G}$ decreases uniformly from $+\infty$
to $-\infty$, while the value of ${\cal G}$ increases
monotonically and analytically. The value of $\Phi_{\cal G} (0)$
is given in (\ref{rest}).
Similar results hold for $R$, which decreases monotonically from
region C to A to B. In the body of this paper, we shall regard
$R$ and ${\cal G}$ as known functions of $r$ and $T$ which can be
looked up in Appendix~\ref{ru} and I.
\newline
({\em ii\/}) Then, we study ${\cal Z}$ directly in $d=2$. We do this mostly with our
own convenient choice of a (virtual) lattice regularization, but we will be
careful to express all physical response functions in terms of $R$
and ${\cal G}$ after using (\ref{defR}); we will explicitly show that our results become
independent of $a$, for small $a$, when this is done.
Finally, for these couplings $R$ and ${\cal G}$, we use the ${\cal Z}_Q$
imposed values given in Appendix~\ref{ru} and (\ref{rest}),
to determine the true physical
response functions. The values of $R$ and ${\cal G}$ vary non-trivially
as a function of $r$ and $T$, and the final results then contain
the crossovers between the different regions of Fig~\ref{fig1}.

Simple, engineering dimensional analysis shows that the
static susceptibility of ${\cal Z}$, and therefore also of ${\cal
Z}_Q$, has the form (as shown in I)
\begin{equation}
T \chi(k) = \frac{1}{n} \sum_{\alpha=1}^n
\left\langle \Phi_{\alpha} (k) \Phi_{\alpha} (-k)
\right\rangle = \frac{T}{R} \Psi \left( \frac{k}{R^{1/2}}, {\cal G}
\right),
\label{defpsi}
\end{equation}
where $\Psi (\overline{k}, {\cal G})$  is a universal
function of its two arguments.
One of our primary tasks here shall be the computation of this
universal function directly
in $d=2$. For small ${\cal G}$, this can be done in
naive perturbation theory; as is clear from (\ref{rest}) such a perturbation theory
is applicable in region A for small $\epsilon$, as ${\cal G} \sim
\sqrt{\epsilon}$. However, in $d=2$ and $\epsilon = 1$, the value
of ${\cal G}$ in (\ref{rest}) is not particularly small, and the
resulting perturbation theory for the static susceptibility is not
quantitatively reliable.
We shall present the results of
straightforward numerical simulations carried out on a small
workstation, which give a reasonably accurate determination of the universal
function $\Psi$ in $d=2$, except when ${\cal G}$ is extremely large.
Somewhat remarkably, precisely in $d=2$, we shall also be
able to make exact statements in the strong coupling limit ${\cal G} \rightarrow
\infty$ through a duality-like transformation, and this will
provide a useful supplement to the numerical results. In summary,
by a combination of weak-coupling perturbation theory, an exact
strong-coupling `duality' mapping, and numerical simulations, we
shall obtain fairly complete knowledge of $\Psi$ directly in
$d=2$.

In $d=2$, for the cases
$n=1,2$, there are critical values ${\cal G} = {\cal G}_c$ where ${\cal Z}$ exhibits
phase transitions (in the universality classes of  $d=2$
classical Ising and Kosterlitz-Thouless respectively),
which appear as singularities of the function $\Psi$ at ${\cal G} =
{\cal G}_c$.
The values of ${\cal G}_c$ will be determined numerically (for $n=1$,
${\cal G}_c$
was obtained in Ref~\onlinecite{loinaz}). For ${\cal Z}_Q$, these
phase transitions, of course, reflect those along the $T>0$ full line
within region B in Fig~\ref{fig1}.

An important property characterizing the low temperature phase
for $d=2$, $n=2$ and ${\cal G} > {\cal G}_c$ is the spin
stiffness, $\rho_s (T)$.
Simple dimensional arguments similar to those leading to (\ref{defpsi})
now show that its temperature dependence obeys
\begin{equation}
\rho_s (T) = T \widetilde{\Psi}_{\rho} ( {\cal G})
\label{n2}
\end{equation}
where $\widetilde{\Psi}_{\rho}$ is a universal scaling
function; it is closely related to the experimentally measurable $\Psi_{\rho}$
in (\ref{r1}), and this will be discussed in Section~\ref{expts}.
Clearly, $\widetilde{\Psi}_{\rho}$ vanishes for ${\cal G} < {\cal
G}_c$.

\subsection{Dynamics}
\label{dynamics}
As we have emphasized in I and II, the dynamical properties of
region A of Fig~\ref{fig1} in $d=2$ are especially interesting because they
are characterized by a phase coherence time, and an inelastic scattering
time,
which are both universal numbers times $ T $. Consequently,  thermal and quantum
fluctuations play equal roles in the dynamical theory; this novel regime of
dynamics was dubbed {\em quantum relaxational} in Ref~\cite{CSY}.
Here we argue that
for the case where $\epsilon$ is small, and for the long wavelength relaxational
dynamics of the order parameter {\em only\/}, it is possible to
disentangle the quantum and classical thermal effects. The
central reason for this is that the predominant modes contributing
to the relaxation of the order parameter fluctuations at long
wavelengths have an energy of order
\begin{equation}
c \sqrt{R} \sim \sqrt{\epsilon} T,
\label{r4}
\end{equation}
from (\ref{rest}). It must be emphasized that these modes carry
negligible amounts of current, and the transport properties
continue to be dominated by excitations with energy of order $T$
even for small $\epsilon$, as we have discussed in II.
As the energy in (\ref{r4}) is parametrically smaller than $T$,
the occupation number of the typical order-parameter modes
is
\begin{equation}
\frac{1}{e^{c\sqrt{R}/T} - 1} \approx \frac{T}{c\sqrt{R}} \sim
\frac{1}{\sqrt{\epsilon}} \gg 1;
\label{r4z}
\end{equation}
The second term above is the classical equipartition value.
So we
can conclude that there is an effective classical {\em non-linear wave
model}
which describes the long wavelength relaxation of the $\phi_{\alpha}$
fluctuations. Quantum effects then appear only in determining the
coupling constants of this effective classical dynamics.

So what is the classical wave model describing the relaxation of the
order parameter modes~? To leading order in $\epsilon$, the
required model can be deduced immediately from some simple general
arguments. First, we have the important constraint that the equal
time correlations must be identical to those implied by ${\cal Z}$
in (\ref{cal}). Second, to endow the $\Phi_{\alpha}$ field with an
equation of motion, we clearly need to introduce a conjugate
momentum variable $\Pi_{\alpha}$. The kinetic energy associated
with this momentum is clearly given by the time derivative term in
${\cal L}_Q$ in (\ref{calsq}). Furthermore, we know that the
co-efficient of this gradient term is not renormalized at order $\epsilon$
when the high frequency modes are integrated out. So we assert
that the required dynamical model is specified by the following partition
function over a classical phase space
\begin{eqnarray}
&& {\cal Z}_C = \int {\cal D} \Phi_{\alpha} (x) {\cal D} \Pi_{\alpha} (x)
\exp \left( - \frac{{\cal H}}{T}
\right) \nonumber \\
&& {\cal H} =  \int d^d x \, \left\{\frac{1}{2} \left[ c^2 \Pi_{\alpha}^2 +
(\nabla_x \Phi_{\alpha})^2 +
\widetilde{R} \, \Phi_{\alpha}^2 \right]
+ \frac{U}{4!} \left( \Phi_{\alpha}^2 \right)^2 \right\}.
\label{calc}
\end{eqnarray}
Notice that we can perform the Gaussian integral over $\Pi_{\alpha}$ exactly,
and we are
left then with the original co-ordinate space in (\ref{cal}): this is the
usual situation in classical statistical mechanics, where momenta
decouple from the static analysis. To
complete the specification of the classical dynamical model, we
need to supplement (\ref{calc}) with equations of motion. These
are obtained simply be replacing the quantum commutators
associated with classical Poisson brackets. So we have
\begin{equation}
\left\{ \Phi_{\alpha} (x) , \Pi_{\beta} (x') \right\}_{P.B.} =
\delta_{\alpha\beta} \delta(x-x').
\label{r5}
\end{equation}
The deterministic, real time equations of motion are then the Hamilton-Jacobi
equations of the Hamiltonian ${\cal H}$, and they are given by
\begin{eqnarray}
\frac{\partial \Phi_{\alpha}}{\partial t} &=&
\left\{ \Phi_{\alpha} (x) , {\cal H} \right\}_{P.B.} \nonumber \\
&=& c^2 \Pi_{\alpha},
\label{r6}
\end{eqnarray}
and
\begin{eqnarray}
\frac{\partial \Pi_{\alpha}}{\partial t} &=&
\left\{ \Pi_{\alpha} (x) , {\cal H} \right\}_{P.B.} \nonumber \\
&=&  \nabla_x^2 \Phi_{\alpha} - \widetilde{R}
\Phi_{\alpha} - \frac{U}{6} (\Phi_{\beta}^2) \Phi_{\alpha},
\label{r7}
\end{eqnarray}
The equations (\ref{calc}), (\ref{r6}) and (\ref{r7}) define the
central dynamical non-linear wave model of interest in this paper. We are
interested in the correlations of the field $\Phi_{\alpha}$ at
unequal times, averaged over the set of initial conditions
specified by (\ref{calc}).
Notice all the thermal `noise' arises only in the random set of
initial conditions. The subsequent time evolution obeys Hamiltonian
dynamics, is
completely deterministic, and precisely conserves energy,
momentum, and total ${\rm O} (n)$ charge. This should be
contrasted with the classical dynamical models studied in the
theory of dynamic critical phenomena~\cite{hhm,halphoh}, where there
are statistical noise terms and an explicit damping coefficient
in the equations of motion.

The dynamical model above has been defined in the continuum,
and so we need to consider the nature of its short distance
singularities. Our primary assertion is that for $d<3$, the {\em only\/}
short distance singularities are those already present in the
equal time correlations analyzed in Section~\ref{statics}.
These were removed by the simple renormalization in (\ref{defR}),
and we maintain this is also sufficient to define the continuum
limit of the unequal time correlations. These assertions rely on
our experience with the structure of the perturbation theory in
${\cal G}$ presented in I, and on the consistency of the numerical data we shall present
with the scaling structure we describe below.

Assuming that introducing $R$ as in (\ref{defR}) allows to take
the limit $a \rightarrow 0$, we can deduce the scaling form of unequal
time correlations by simple dimensional arguments.
We define the dynamic structure factor, $S(k,
\omega)$ by
\begin{equation}
S (k, \omega) =  \frac{1}{n} \int_{-\infty}^{\infty} d t \int d^d x
\,
\sum_{\alpha=1}^{n}
\langle \Phi_{\alpha} (x, t) \Phi_{\alpha} (0,0) \rangle e^{-i(kx - \omega
t)}.
\label{defsko}
\end{equation}
Notice that, unlike (\ref{defchi}), this involves an integral over
real time, $t$. Comparing with the equal time correlator in (\ref{defpsi}),
we clearly have the relation
\begin{equation}
T \chi (k) = \int_{-\infty}^{\infty} \frac{d \omega}{(2 \pi)} S(k,
\omega).
\label{r8}
\end{equation}
However, what is the relationship between $S(k, \omega)$ and the
physically appropriate quantum dynamic susceptibility $\chi (k, \omega)$ obtained by
analytically continuing (\ref{defchi}) ? By analogy with
(\ref{defsko}) we can define the physical, quantum dynamic structure factor
\begin{equation}
S_Q (k, \omega) =  \frac{1}{n} \int_{-\infty}^{\infty} d t \int d^d x
\,
\sum_{\alpha=1}^{n}
\langle \phi_{\alpha} (x, t) \phi_{\alpha} (0,0) \rangle e^{-i(kx - \omega
t)},
\label{defqsko}
\end{equation}
where it is understood the time evolution is now due to the
quantum Hamiltonian implied by ${\cal Z}_Q$, and so is the thermal
average. This structure factor obeys an exact fluctuation
dissipation relation to $\chi (k, \omega)$ defined in
(\ref{defchi}):
\begin{equation}
S_Q (k, \omega) = \frac{2}{1 - e^{-\omega/T}} \mbox{Im} \chi (k,
\omega).
\label{fdt}
\end{equation}
We can relate the dynamic structure factors $S_Q$ and $S$
only under the conditions that the dominant spectral weight of
excitations is at an energy smaller than $T$. As argued earlier,
this is the case here by (\ref{r4}) for $\epsilon$ small. Assuming
this condition we have $S(k,\omega) \approx S_Q (k, \omega)$
and
\begin{equation}
S (k, \omega) \approx \frac{2T}{\omega} \mbox{Im} \chi (k,
\omega).
\label{cfdt}
\end{equation}
Finally, we can write down the scaling form obeyed by $S(k,
\omega)$;
from the arguments of the previous paragraph, and simple
engineering dimensional considerations as in (\ref{defpsi}), we obtain
\begin{equation}
S(k, \omega) = \frac{T \chi(k) }{c R^{1/2}} \Psi_{Sc} \left(
\frac{k}{\sqrt{R}}, \frac{\omega}{c \sqrt{R}}, {\cal G} \right),
\label{r9}
\end{equation}
where $\Psi_{Sc} ( \overline{k}, \overline{\omega}, {\cal G})$ is a
dimensional universal function is an even function of $\overline{\omega}$.
The prefactor of (\ref{r9}) has been chosen to ensure that this
function has a constant integral over frequency
\begin{equation}
\int_{-\infty}^{\infty} \frac{d \overline{\omega}}{2 \pi}
\Psi_{Sc} ( \overline{k}, \overline{\omega}, {\cal G}) = 1,
\label{r10}
\end{equation}
as follows immediately from
(\ref{r8}), or from (\ref{cfdt}) after use of the Kramers-Kronig
representation of $\chi(k)$ in terms of $\mbox{Im} \chi (k,
\omega)$.
We will obtain information on the structure of $\Psi_{Sc}$
for $d=2$ in Section~\ref{dyn}. We shall be
especially in interested in the $\omega \rightarrow 0$
limit of $S(k, \omega)$ which describes the long time correlations of the
order parameter $\Phi_{\alpha}$; this limit is not accessible in
perturbation theory even for small ${\cal G}$, as was shown in I.

The outline of the remainder of the paper is as follows. We will
present details of our analysis of the static properties of ${\cal Z}$
in Section~\ref{stat}. The main achievements of this section are
of a technical and quantitative nature, and there are no
qualitatively new physical results; some readers may wish to skip
this section and go directly to Section~\ref{dyn}. In
Section~\ref{dyn} we will present our numerical results on the
long time dynamics of the model ${\cal Z}_C$. A synthesis of our
results in the context of its experimental implications will then
appear in Section~\ref{expts}. Some technical details, including a
summary of needed results from I and II are presented in the
Appendices.

\section{Statics in two dimensions}
\label{stat}
This section will examine the classical model (\ref{cal}) directly
in $d=2$. We will obtain essentially complete information on the
static susceptibility, $\chi (k)$, by a combination of weak
coupling (Section~\ref{statweak}), strong coupling
(Section~\ref{statstrong}), and numerical methods
(Section~\ref{num}). The exact duality-like transformation will be
described in Section~\ref{statstrong}.

\subsection{Weak coupling}
\label{statweak}
For small ${\cal G}$, we can perform a familiar Feynman graph
expansion in the quartic coupling in ${\cal Z}$. At order
${\cal G}^2$ we obtain at $k=0$
\begin{equation}
\chi^{-1} (0) = R \left[ 1 -  \left(\frac{n+2}{18}\right) J(1,1,1) {\cal G}^2 + {\cal
O} ({\cal G}^3) \right],
\label{w1}
\end{equation}
where $J(1,1,1)$ is a number defined in (\ref{dual5i}), and
(\ref{w1}) is clearly consistent with the scaling form
(\ref{defpsi}).
The order ${\cal G}^2$ correction starts becoming important for ${\cal G} \approx
15$, which then is roughly where the crossover to strong coupling
occurs.

The spatial correlations of $\Phi$ decay exponentially on a scale
of order $1/\sqrt{R}$. Neglecting the order ${\cal G}^2$
corrections, we get from a Fourier transform of the $1/(k^2 + R)$
propagator at large $|x|$
\begin{equation}
\frac{1}{n}\sum_{\alpha=1}^{n} \langle \Phi_{\alpha} (x) \Phi_{\alpha} (0) \rangle =
\frac{T}{\sqrt{8 \pi |x| /\xi}} e^{-|x|/\xi},
\label{w2}
\end{equation}
where the correlation length $\xi$ is
\begin{equation}
\xi = \frac{1}{\sqrt{R}} \left[ 1 + {\cal O} ({\cal G}^2) \right].
\label{w3}
\end{equation}

\subsection{Strong coupling}
\label{statstrong}
We will now consider correlators of ${\cal Z}$ in the limit ${\cal G} \rightarrow
\infty$. Quite remarkably, exact information can be obtained in
this limit too. The key is an ingenious proposal made some time
ago by Chang~\cite{chang} for the $n=1$ Ising case, but which
appears to have been forgotten since. Chang proposed a
strong-to-weak coupling mapping for $n=1$, which has the flavor to a duality
transformation. Here we will review his mapping, and show that
closely related methods can be applied to all $n$.

The argument begins by noting that in the limit $R \rightarrow 0$,
the bare mass $\widetilde{R}$ in (\ref{defR}) tends to $-\infty$.
So at short scales, the effective potential controlling the
fluctuations of $|\Phi_{\alpha}|$ will have a negative curvature
at the origin and a minimum at a non-zero value of
$|\Phi_{\alpha}|$. This suggests that we renormalize the theory
with a {\em negative} renormalized co-efficient of the $\Phi_{\alpha}^2$
term. So we replace (\ref{defR}) by
\begin{equation}
\widetilde{R} = -\frac{R_D}{2} - T U \left(\frac{n+2}{6} \right) \int^{1/a}
\frac{d^2 k}{(2 \pi)^2} \frac{1}{k^2 + R_D},
\label{defQ}
\end{equation}
where we have introduced a new renormalized `dual' mass $R_D>0$; the factor
of 1/2 in the first term on the right hand side is for future
convenience, and has no particular significance. While both
(\ref{defQ}) and (\ref{defR}) have a non-universal
cutoff-dependence in their momentum integral, this disappears when
we combine them to eliminate $\widetilde{R}$ and obtain
\begin{equation}
R + \frac{R_D}{2} = T U \left(\frac{n+2}{24 \pi} \right) \ln
\frac{R_D}{R}.
\label{s1}
\end{equation}
This equation can be solved to yield $R_D>0$ as a function of $U$
and $R$. As we will discuss shortly, such a solution exists only
for $R$ small enough.

\begin{figure}
\epsfxsize=6in
\centerline{\epsffile{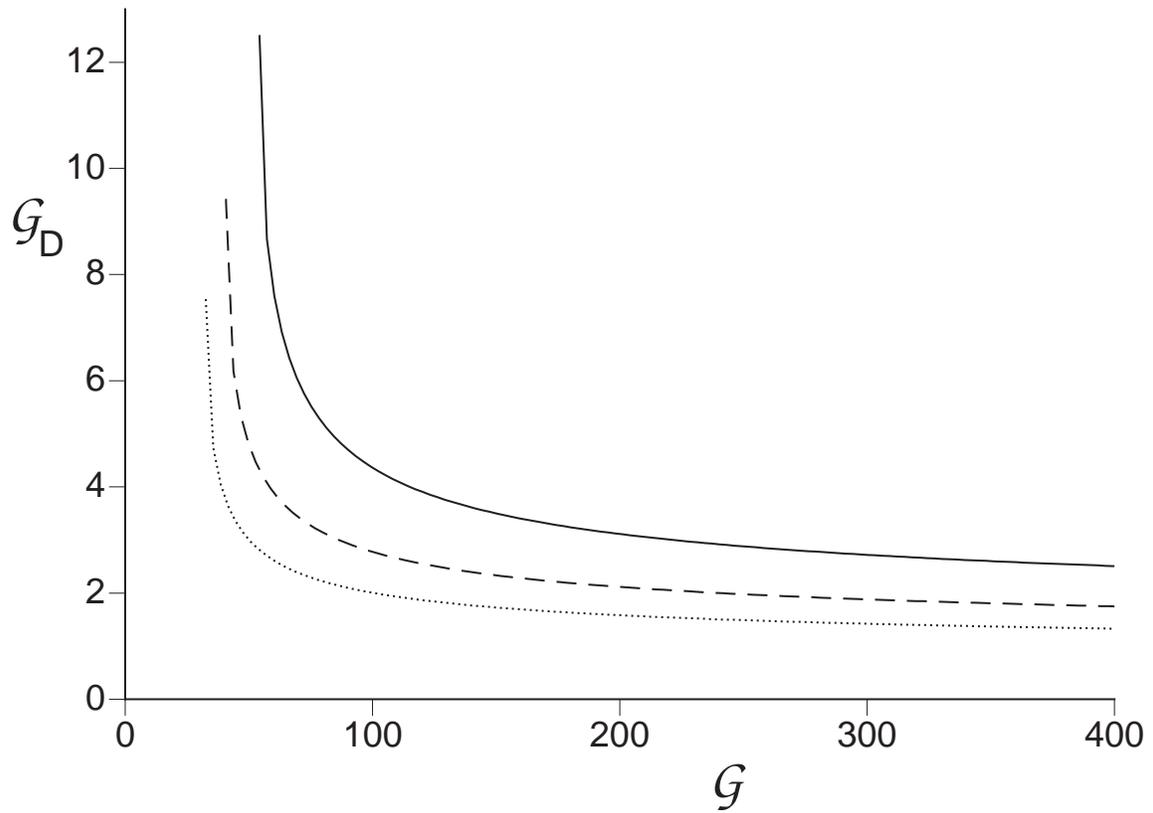}}
\caption{The `dual' dimensionless coupling ${\cal G}_D$ as a function
of ${\cal G}$ for $n=1$ (full line), $n=2$ (dashed line),
and $n=3$ (dotted line).}
\label{fig3}
\end{figure}

Assuming the existence of a solution (\ref{s1}) for $R_D$, we now have a new
renormalized theory in which the local effective potential for
$\Phi_{\alpha}$ fluctuations has the form
\begin{equation}
- \frac{R_D}{4} \Phi_{\alpha}^2 + \frac{U}{4\!} ( \Phi_{\alpha}^2 )^2
+ \ldots,
\label{s2}
\end{equation}
where the ellipses represent counter-terms arising from
(\ref{defQ}) which will cancel the cutoff dependencies order by
order in $U$.
So if this renormalized theory is weakly coupled, $|\Phi_{\alpha}|$
will fluctuate around a non-zero mean value of order $\sqrt{3
R_D/U}$. But is this the case~?
Clearly, this depends upon the value of a `dual' dimensionless
coupling, ${\cal G}_D$, which is the analog of (\ref{deflam})
\begin{equation}
{\cal G}_D = \frac{T U}{R_D}.
\label{s3}
\end{equation}
What is the value of ${\cal G}_D$~? This can be obtained by
combining (\ref{s1}), (\ref{deflam}) and (\ref{s3}) into the
dimensionless equation
\begin{equation}
\frac{1}{{\cal G}} + \frac{1}{2 {\cal G}_D} = \left( \frac{n+2}{24
\pi} \right) \ln \frac{{\cal G}}{{\cal G}_D}
\label{s4}
\end{equation}
which can be solved to yield ${\cal G}_D$ as a function of
${\cal G}$. A plot of ${\cal G}_D$ versus ${\cal G}$ is shown in
Fig~\ref{fig3}.
A straightforward analysis of (\ref{s4}) shows that there is no
real solution for ${\cal G}_D$ for ${\cal G} < 54.2756$ for $n=1$,
for ${\cal G} < 40.7069$ for $n=2$, and ${\cal G} < 32.5655$
for $n=3$. For ${\cal G}$ larger than these values, there is a
solution for ${\cal G}_D$ which decreases monotonically from
$({\cal G}_D)_{\rm max} = 12 \pi/(n+2)$ as ${\cal G}$ increases,
and as ${\cal G} \rightarrow \infty$ it obeys
\begin{equation}
{\cal G}_D \approx \frac{24 \pi}{(n+2)} \frac{1}{\ln {\cal G}}.
\label{s5}
\end{equation}
So ${\cal G}_D$ is small as ${\cal G}$ becomes very large, and this
`dual' problem is therefore in a weak coupling limit, as we had
hoped.

We have so far been expressing all universal physical properties of
(\ref{cal}) in terms of ${\cal G}$ and $R$. However, for large ${\cal
G}$, it is clear that we can freely trade these couplings for ${\cal G}_D$
and $R_D$: we define ${\cal G}_D$ from (\ref{s4}) and relate $R_D$ and $R$
by the relationship
\begin{equation}
R_D {\cal G}_D  = R {\cal G},
\label{s5a}
\end{equation}
obtained by comparing (\ref{deflam}) and (\ref{s3}).
We will mainly use $R_D$ and ${\cal G}_D$ as our independent couplings
in the remainder of this
subsection.

It now remains to do a weak coupling analysis
in powers of ${\cal G}_D$. This is straightforward for $n=1$, but the
cases with continuous symmetry, $n\geq 2$, have to be treated with
some care. All of these analyses have been discussed in
Appendix~\ref{dual}, and we will present the final results for different value
of $n$ in the following subsections.

\subsubsection{$n=1$}
\label{ising}

For $n=1$, we have from (\ref{dual4}) for large $|x|$
\begin{equation}
\langle \Phi_{\alpha} (x) \Phi_{\alpha} (0) \rangle = N_0^2 +
\frac{T}{\sqrt{8 \pi |x| /\xi}} e^{-|x|/\xi},
\label{s6}
\end{equation}
where the correlation length
\begin{equation}
\xi = \frac{1}{\sqrt{R_D}},
\label{s7}
\end{equation}
and the spontaneous magnetization was computed in (\ref{dual4c}):
\begin{equation}
N_0 = \sqrt{\frac{3T}{{\cal G}_D}} \left[
1 - \frac{5 {\cal G}_D^2}{12} J(1,1,1) + {\cal O} ( {\cal G}_D^3 )
\right].
\label{s7aa}
\end{equation}
The numerical constant $J(1,1,1)$ also appeared in (\ref{w1}), and
its value is given in (\ref{dual5i}).
So the correlator approaches $N_0^2$ exponentially on the scale
$\xi$.
In momentum space, the static susceptibility has the form
\begin{equation}
\chi (k) = \frac{N_0^2}{T} (2 \pi)^2  \delta^2 (k) +
\frac{1}{k^2 + R_D},
\label{s7a}
\end{equation}
which is of the form (\ref{defpsi}).
The presence of the delta function indicates true
long-range order which breaks the $\Phi \rightarrow -\Phi$
symmetry for small ${\cal G}_D$ (large ${\cal G}$).

\subsubsection{$n=2$}
\label{xy}

For $n=2$,
from (\ref{dual5}) and the arguments below it, we can conclude
that for $|x| \gg 1/\sqrt{R_D}$, there is a power-law decay in the order parameter
correlator
\begin{equation}
\sum_{\alpha=1}^{n} \langle \Phi_{\alpha} (x) \Phi_{\alpha} (0) \rangle = \frac{3
T}{{\cal G}_D} \left( 1 + \frac{(\ln2 - \gamma)}{6 \pi} {\cal G}_D + {\cal O}
({\cal G}_D^2) \right) \left[ |x| \sqrt{R_D} \right]^{-\eta},
\label{s8}
\end{equation}
where $\gamma$ is Euler's constant, and
the continuously varying exponent, $\eta$, is related to
$\rho_s (T) $, the exact renormalized spin stiffness towards $O(2)$ rotations, by
\begin{equation}
\eta = \frac{T}{2 \pi \rho_s (T)}.
\label{s8a}
\end{equation}
We computed $\rho_s (T)$
in Appendix~\ref{dualxy} in a perturbation theory in ${\cal G}_D$
and found
\begin{equation}
\frac{\rho_s (T)}{T} = \frac{3}{{\cal G}_D} - \frac{{\cal G}_D}{36} + {\cal O} ({\cal
G}_D^2),
\label{s9}
\end{equation}
which is consistent with the scaling form (\ref{n2}).
The power-law decay in (\ref{s8}) corresponds to a quasi-long-range XY order in
the two-component planar order parameter $(\Phi_1, \Phi_2)$.
In momentum space, the quasi-long range order implies a power-law
singularity in the static susceptibility at $k=0$:
\begin{equation}
\chi (k) = \frac{3 \Gamma (1-\eta/2)}{\pi 2^{\eta} \Gamma(\eta/2)}
\frac{1}{{\cal G}_D R_D^{\eta/2} k^{2-\eta}},
\label{s9a}
\end{equation}
which is consistent with the scaling form (\ref{defpsi}).

\subsubsection{$n \geq 3$}
\label{heisenberg}

For $n \geq 3$, no long-range or quasi-long-range order is possible.
Correlations always decay exponentially at sufficiently long
scales, but the correlation length does become very large in the
strong coupling limit.
We conclude from the analysis in
Section~\ref{dualheis} that the ultimate long-distance decay of
the correlation function has the form
\begin{equation}
\sum_{\alpha=1}^n \langle \Phi_{\alpha} (x) \Phi_{\alpha} (0) \rangle = {\cal C}_1 \frac{
T}{{\cal G}_D} {\cal G}_D^{(n-1)/(n-2)} \frac{e^{-|x|/\xi}}{\sqrt{|x|/\xi}},
\label{s10}
\end{equation}
where ${\cal C}_1$ is a universal number, and the correlation length is
given by
\begin{equation}
\xi = \frac{1}{\sqrt{R_D}} \Gamma\left(\frac{n-1}{n-2} \right)
\left[\frac{e(n-2) {\cal G}_D}{48 \pi} \right]^{1/(n-2)} \exp\left(
 \frac{6
\pi}{(n-2){\cal G}_D} \right)
\label{s11}
\end{equation}
The static susceptibility can be deduced from the results of
Ref~\cite{CHN,CSY}, and is given by
\begin{equation}
\chi (k) = {\cal C}_2 \left( \frac{{\cal G}_D}{3} \right)^{1/(n-2)} \xi^2 f(k \xi)
\label{s11a}
\end{equation}
for small $k$, where ${\cal C}_2$ is a universal number (${\cal C}_2 \approx 1.06$
for $N=3$~\cite{tyc}), $f$ is a smooth scaling function considered in
Refs~\cite{CHN,CSY} with $f(0)=1$. Notice that, unlike the cases $n=1,2$, there
is no singularity in $\chi (k)$ at $k=0$ for small ${\cal G}_D$ (large ${\cal
G}$); instead $\chi (0)$ becomes exponentially large ($\sim
\xi^2$), has an exponentially small width ($\sim 1/\xi$) in
momentum space, but remains a smooth function of $k$.

\subsection{Numerical results}
\label{num}

We now have an understanding of the properties of ${\cal Z}$ both
in the limits ${\cal G} \rightarrow 0$ and ${\cal G} \rightarrow \infty$
in $d=2$. For small ${\cal G}$, we have the result
(\ref{w2},\ref{w3}) showing the $\Phi_{\alpha}$ correlations decay
exponentially in space due to the fluctuations of $n$ $\Phi_{\alpha}$ modes about
$\Phi_{\alpha} = 0$, while (\ref{w1}) shows that the static
susceptibility $\chi (0) \sim 1/R$. For large ${\cal G}$ we have
the results for the static susceptibility in (\ref{s7a}),
(\ref{s9a}), and (\ref{s11a}). We will examine the manner in
which the system interpolates between these limits in the
following subsections.

\subsubsection{$n=1$}
\label{num1}

The delta function in (\ref{s7a}) indicates the presence of
long-range order for sufficiently large ${\cal G}$. This delta
function is expected first appears at a critical value ${\cal G}={\cal G}_c$
by a phase transition in the universality class of the $d=2$
classical Ising model. The value of ${\cal G}_c$ was determined
numerically in a recent Monte Carlo simulation in
Ref~\cite{loinaz}, which found
\begin{equation}
{\cal G}_c = 61.44.
\label{n1}
\end{equation}

\subsubsection{$n=2$}
\label{num2}
The quasi-long range order implied by (\ref{s8}) and (\ref{s9a})
is expected to be present for ${\cal G} > {\cal G}_c$,
and to vanish at ${\cal G}_c$ by a Kosterlitz-Thouless transition.
The low temperature phase is characterized by the spin stiffness
which obeys the scaling form in (\ref{n2}). The
behavior of the scaling function $\widetilde{\Psi}_{\rho}$ for large ${\cal G}$
(small ${\cal G}_D$) was obtained earlier in (\ref{s9}). We expect
that
\begin{equation}
\widetilde{\Psi}_{\rho} ({\cal G} < {\cal G}_c) = 0,
\label{n2a}
\end{equation}
while precisely at ${\cal G}_c$ its takes the value specified by
the Nelson-Kosterlitz jump
\begin{equation}
\widetilde{\Psi}_{\rho} ({\cal G} =  {\cal G}_c ) = \frac{2}{\pi}
\label{n3}
\end{equation}

We performed Monte Carlo simulations to obtain more information on
the functional form of $\widetilde{\Psi}_{\rho}$ and the value of
${\cal G}_c$. We discretized (\ref{cal}) on a square lattice of
spacing $a$, and used an $L\times L$ lattice with periodic boundary
conditions. The Monte Carlo sweeps consisted of two alternating
steps. First we updated both the amplitude and phase of $(\Phi_1 +
i \Phi_2)$ on each site by a heat bath algorithm~\cite{toral}. Then
we applied the Wolff cluster algorithm~\cite{wolff} to rotate the
phase of sites on clusters by a random angle. As we are interested
in fairly large values of ${\cal G}$, where
$\widetilde{\Psi}_{\rho}$ is non-zero in the thermodynamic limit,
it is more appropriate to use the dual couplings, ${\cal G}_D$ and
$R_D$ in testing for the appearance of the continuum limit. In
particular, we need $R_D a^2
\ll 1$, and we used values around $R_D a^2 \approx 0.04$; this was
found to yield $a$-independent susceptibilities, as we will display
more explicitly in our discussion of the $n=3$ case.

\begin{figure}
\epsfxsize=6in
\centerline{\epsffile{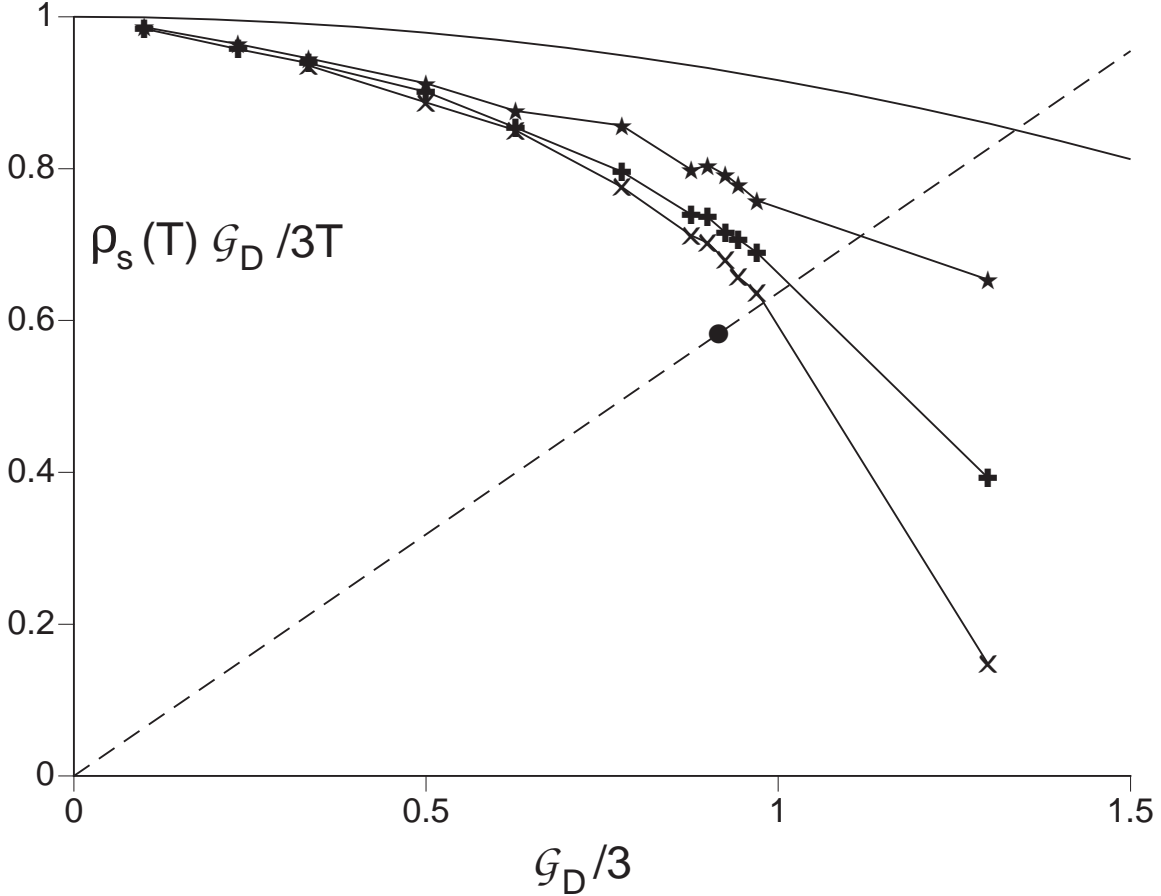}}
\caption{Numerical results for $(\rho_s (T)/T)/({\cal G}_D /3)$ as a function
of ${\cal G}_D /3$. We used a square lattice of $L \times L$ sites
with periodic boundary conditions and lattice spacing $a$.
The lattice sizes used were $L=64$ (stars), $L=128$ (pluses)
and $L=256$ (X's).
The dashed line is the locus of points where the Nelson-Kosterlitz
jump (\protect\ref{n3}) is obeyed. The full line is the result of
the small ${\cal G}_D$ expansion in (\protect\ref{s9}). The filled circle
indicates the position of the Kosterlitz-Thouless transition
determined by the extrapolation to $L = \infty$ limit using the
method described in the text and in Fig~\protect\ref{fig5}.
In the approximation in which we assume that the $T$
dependence of ${\cal G}_D$ is given by its leading value as $T \rightarrow 0$,
${\cal G}_D \approx 3 T/\rho_s (0) $, the scale on the horizontal axis becomes
$T/\rho_s (0)$, while that on the vertical axis becomes $\rho_s (T)/\rho_s (0)$.
}
\label{fig4}
\end{figure}

Our numerical results for $\rho_s (T)$ are shown in Fig~\ref{fig4}.
The stiffness was measured by evaluating the expectation value of
the appropriate current-current correlation function implied by
the Kubo formula. The results are presented by plotting
$(\rho_s (T) /T)/(3/ {\cal G}_D)$ versus ${\cal G}_D/3$, and we
will now discuss the reason for this choice. From (\ref{s9})
and our discussion in Appendix~\ref{ru} we see that ${\cal G}_D$
vanishes linearly as $T \rightarrow 0$, and that
\begin{equation}
\lim_{T \rightarrow 0} \frac{{\cal G}_D ( T )}{T} = \frac{3}{\rho_s
(0)},
\label{n4}
\end{equation}
where we have now emphasized that ${\cal G}_D$ is a function of
$T$, as was also noted in the scaling form (\ref{phig}); this
relationship guarantees that vertical co-ordinate in
Fig~\ref{fig4} becomes unity as ${\cal G}_D \rightarrow 0$.
Further, if we approximate the $T$ dependence of ${\cal G}_D$ by
${\cal G}_D (T) \approx 3 T/\rho_s (0)$ (this relationship is {\em
not} exactly true), then the vertical axis in Fig~\ref{fig4}
becomes $\rho_s (T)/\rho_s (0)$, while the horizontal axis is
$T /\rho_s (0)$.

\begin{figure}
\epsfxsize=6in
\centerline{\epsffile{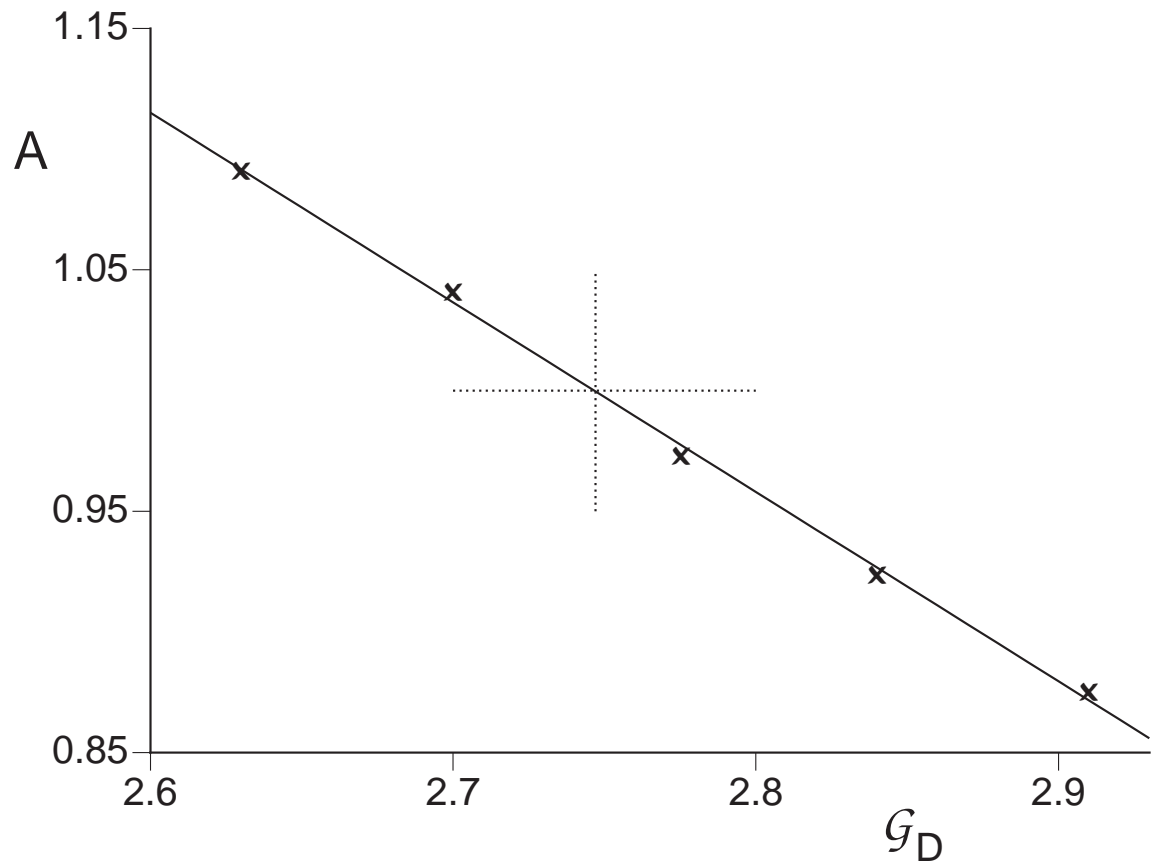}}
\caption{Values of the coefficient $A$ in (\protect\ref{n5}) determined by
fitting the $L$ dependence of the measured $\rho_s (T)/T$ to (\protect\ref{n5}).
We fit the value of $A$ to a linear function of ${\cal G}_D$, and point where the line
has the value $A=1$ (indicated by the dotted lines), determines the position
of the Kosterlitz Thouless transition.
}
\label{fig5}
\end{figure}

Notice that for small ${\cal G}_D$, the results for $\rho_s (T)$
are approximately independent of $L$, while they become
strongly $L$ dependent around ${\cal G}_D \approx 3$, as would be
expected in the vicinity of the Nelson-Kosterlitz jump, which is
present only in the infinite $L$ limit. We can make quite a
precise estimate of the position of this jump by fitting~\cite{QXY} the $L$
dependence of $\rho_s$ to the following theoretically
predicted~\cite{Weber}
finite-size scaling form
\begin{equation}
\frac{\rho_s (T)}{T} = \frac{2A}{\pi} \left( 1 + \frac{1}{2 \ln
(L/L_0)} \right);
\label{n5}
\end{equation}
$A$ and $L_0$ are free parameters, determined by optimizing the
fit. The best fit values of $A$ are shown in Fig~\ref{fig5}.
The value of ${\cal G}_D$ at which $A=1$ determines the position
of the Kosterlitz-Thouless transition, and in this manner we
determine
\begin{equation}
{\cal G}_{Dc} = 2.747.
\label{n6}
\end{equation}
Finally, using (\ref{s4}) we get
\begin{equation}
{\cal G}_c = 102.
\end{equation}

\subsubsection{$n=3$}
\label{num3}

\begin{figure}
\epsfxsize=7in
\centerline{\epsffile{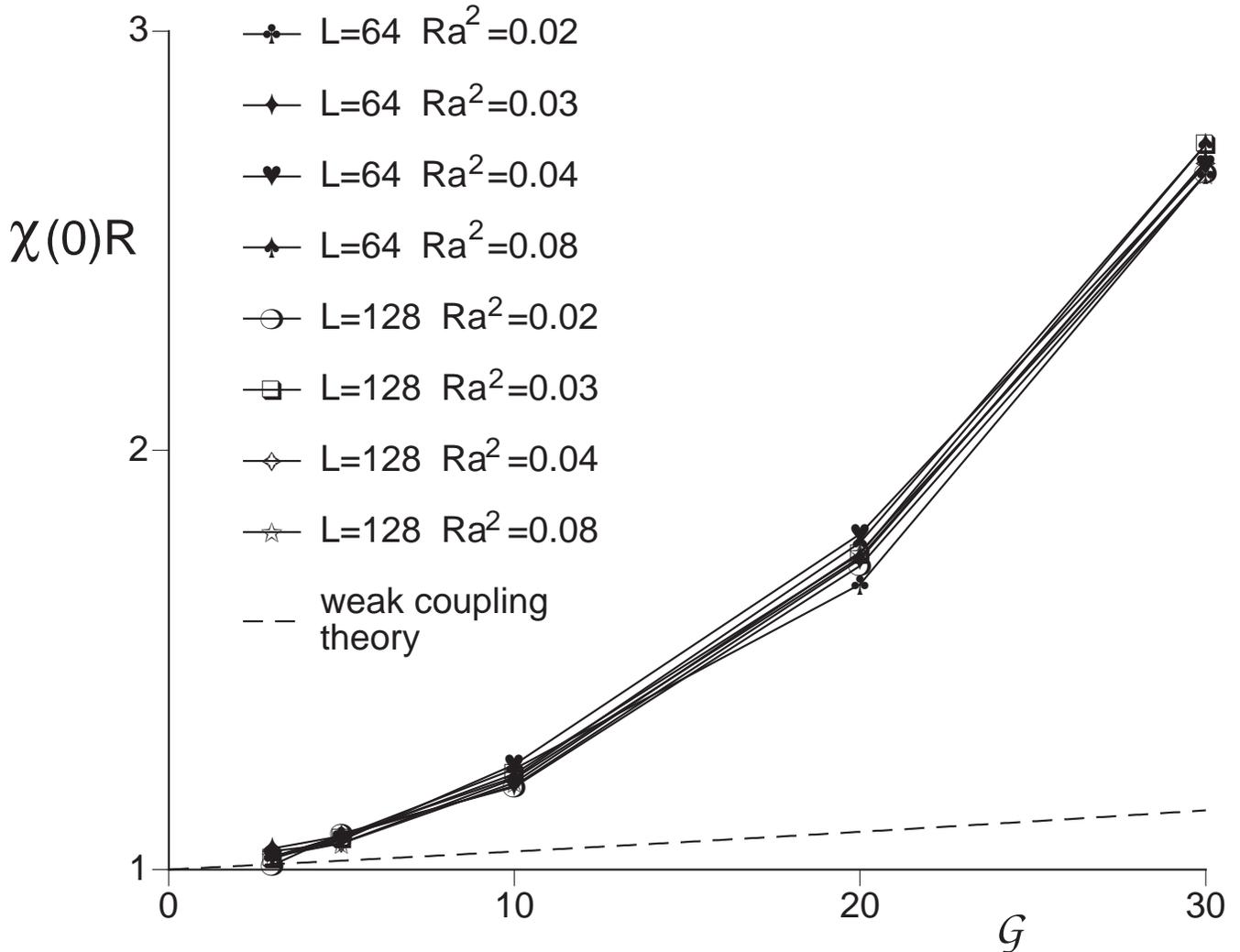}}
\caption{Scaling plot of the static susceptibility $\chi (0)$ for $n=3$
as function of ${\cal G}$.
The dashed line indicates the prediction of weak-coupling expansion
in (\protect\ref{w1}).
}
\label{fig6}
\end{figure}

There is now no phase transition as a function ${\cal G}$, and the
susceptibility exhibits a smooth crossover from the weak-coupling
form (\ref{w1}) to the strong-coupling limit (\ref{s11a}). We
obtained numerical results for $\chi (0)$ at intermediate values
of ${\cal G}$, and the results are shown in Fig~\ref{fig6}.
A range of values of $L$ and $R a^2$ were used, and the excellent
collapse of these measurements in Fig~\ref{fig6} indicates that we
are studying ${\cal L}$ in (\ref{cal}) in the continuum and
infinite volume limits. The weak-coupling prediction of (\ref{w1})
is also shown, and this is seen to work only for very small values
of ${\cal G}$.

\section{Dynamics in two dimensions}
\label{dyn}

\begin{figure}
\epsfxsize=7in
\centerline{\epsffile{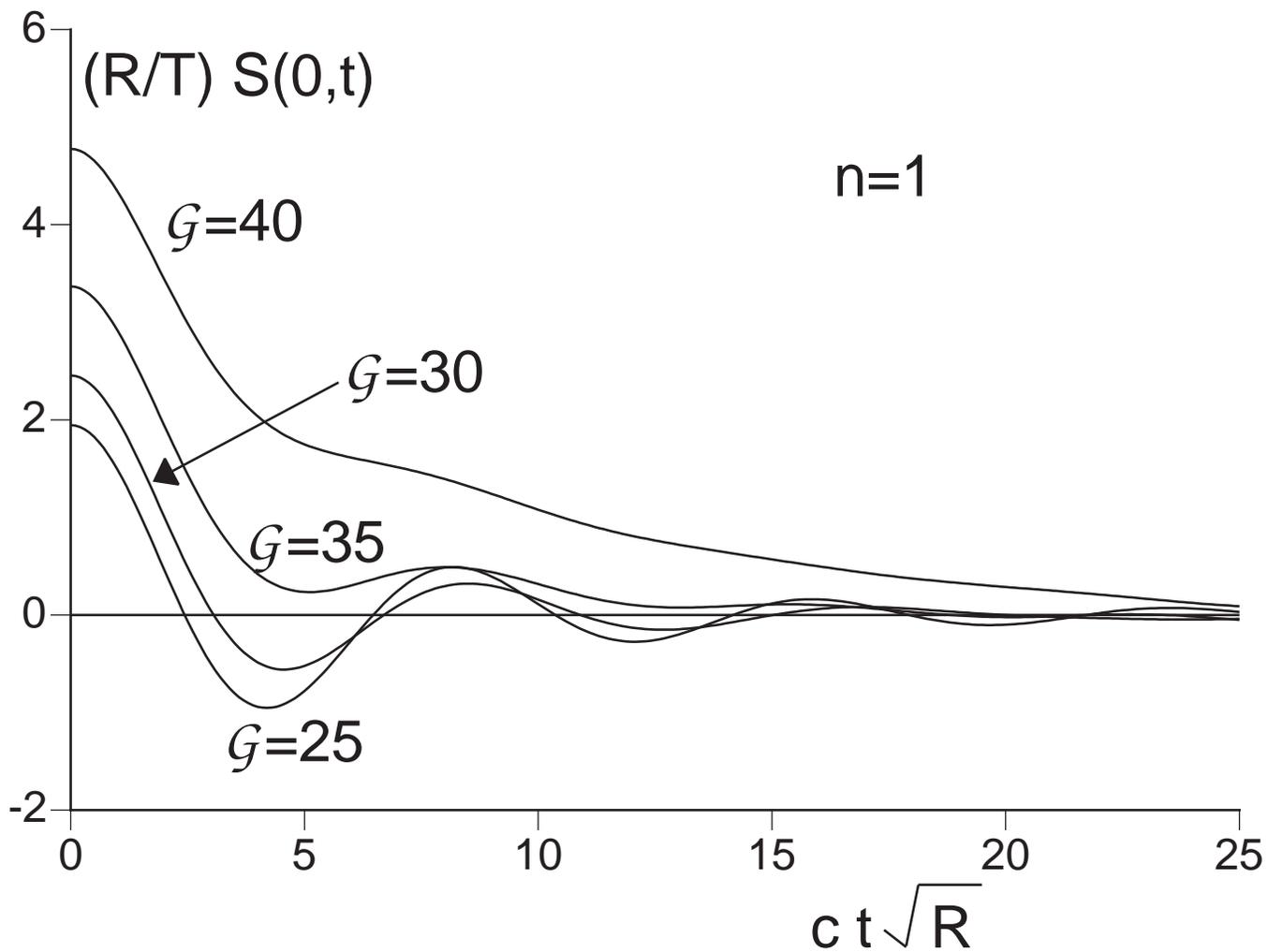}}
\caption{The dynamic structure factor in the time domain,
$S(k,t) = \int d \omega /(2 \pi) S(k, \omega) e^{-i\omega t}$, for
$n=1$ and with ${\cal G} = 25, 30, 35, 40$.
}
\label{fig7}
\end{figure}

\begin{figure}
\epsfxsize=7in
\centerline{\epsffile{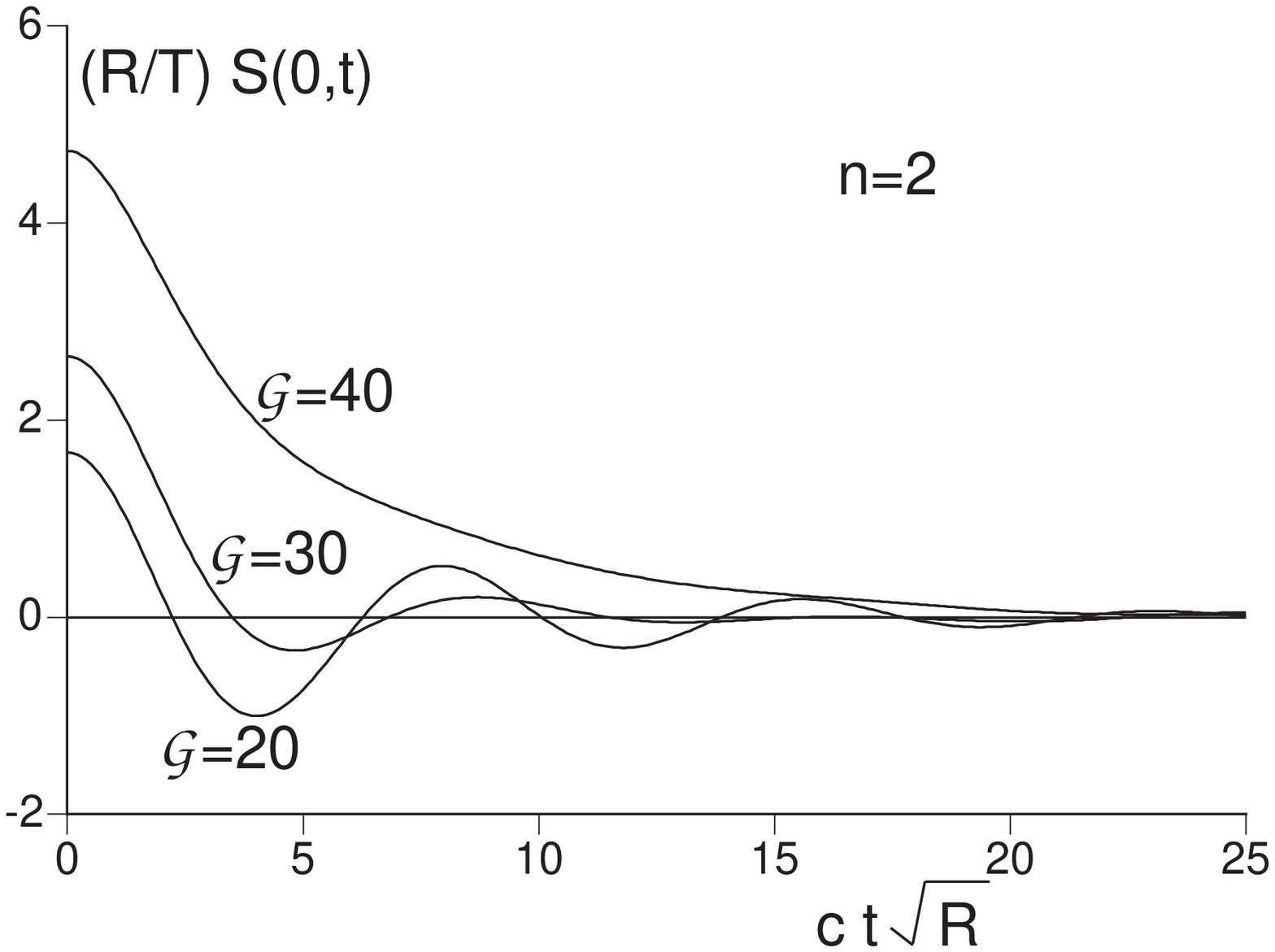}}
\caption{As in Fig~\protect\ref{fig7}, but for
$n=2$ and with ${\cal G} = 20, 30, 40$.
}
\label{fig8}
\end{figure}

\begin{figure}
\epsfxsize=7in
\centerline{\epsffile{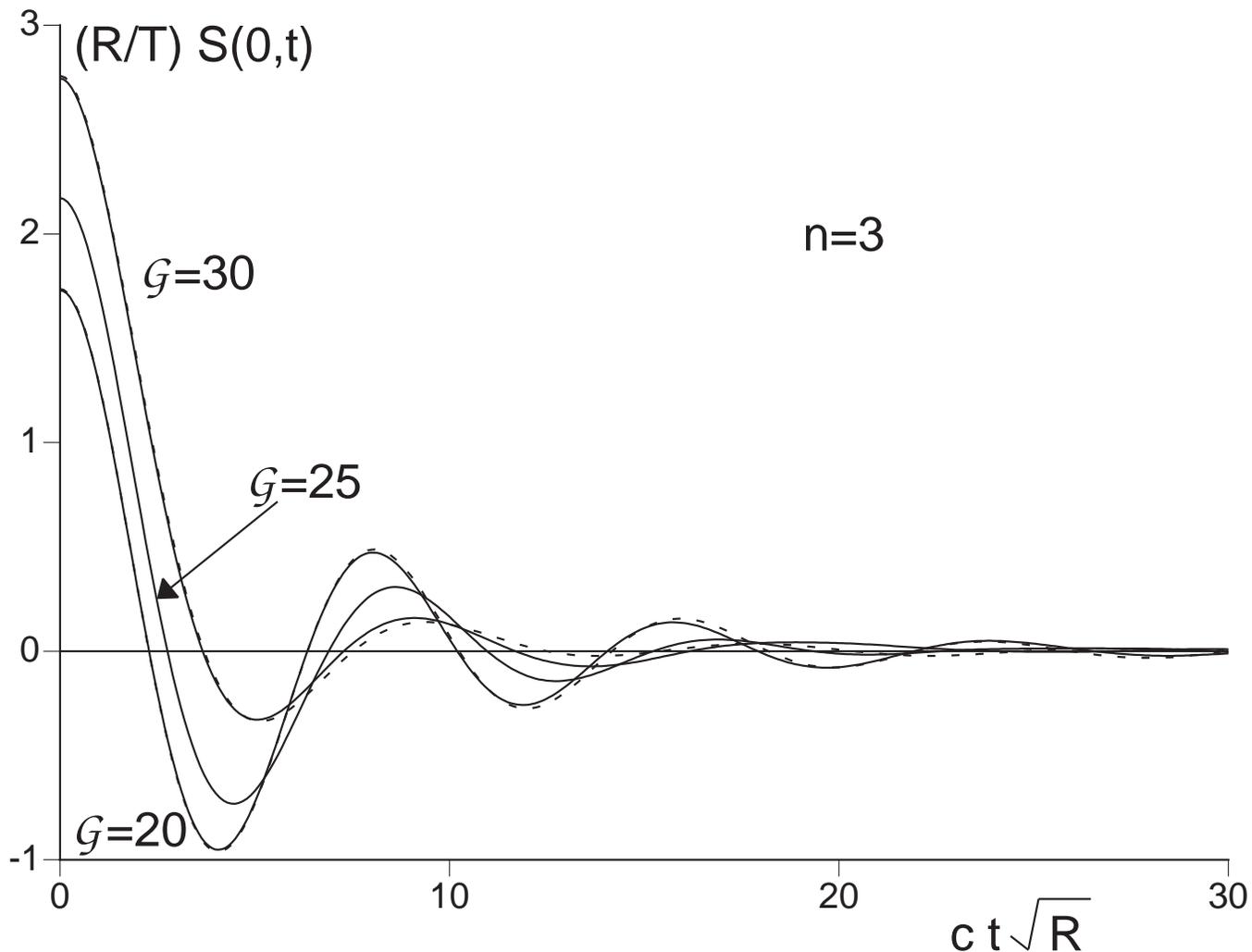}}
\caption{
As in Fig~\protect\ref{fig7} but for $n=3$ and ${\cal G} = 20, 25, 30$.
For ${\cal G}=20,30$ we used two different values of $R a^2$
($Ra^2 = 0.03,0.04$ for ${\cal G}=20$, and $R a^2 = 0.04,0.08$
for ${\cal G}=30$) and these are indicated by the presence of both
dashed and full lines for these cases. The good agreement
between the dashed and full lines is evidence that we are
measuring the universal values in the continuum limit.
}
\label{fig9}
\end{figure}

Finally, we turn to the central problem of dynamic correlations.
We generated initial conditions for $\Phi_{\alpha}$
as described in Section~\ref{num}, for $\Pi_{\alpha}$ by the
simple independent Gaussian distributions specified by
(\ref{calc}),
and then integrated the equations of motion (\ref{r6}) and
(\ref{r7}) by a fourth-order predictor-corrector algorithm.
Correlations of $\Phi_{\alpha}$ at unequal times were then
measured, and in this manner we obtained the correlation function
$\int d^2 x \left\langle \Phi_{\alpha} (x, t) \Phi_{\alpha} (0,0)
\right\rangle$. The results are shown in Figs~\ref{fig7}--\ref{fig9}
for $n=1,2$ and $3$ respectively, for a series of
values of ${\cal G}$.
The values of ${\cal G}$ were chosen to be
around the quantum-critical value (\ref{rest}) evaluated directly
in $\epsilon=1$. Also in Fig~\ref{fig9}, we show results for
different values of $Ra^2$, and their independence on this
parameter is evidence that we are
measuring the universal values in the continuum limit.

There is a simple, and important, trend in the dynamical correlations
with increasing ${\cal G}$. For small ${\cal G}$, the $k=0$
correlations show a clear damped oscillation in time. These
oscillations represent {\em amplitude fluctuations} in $\Phi_{\alpha}$
about a minimum in the effective potential at $\Phi_{\alpha} = 0$.
The damping of the oscillations increases with increasing ${\cal
G}$, until the oscillations disappear entirely for ${\cal G}$
large enough.

\begin{figure}
\epsfxsize=7in
\centerline{\epsffile{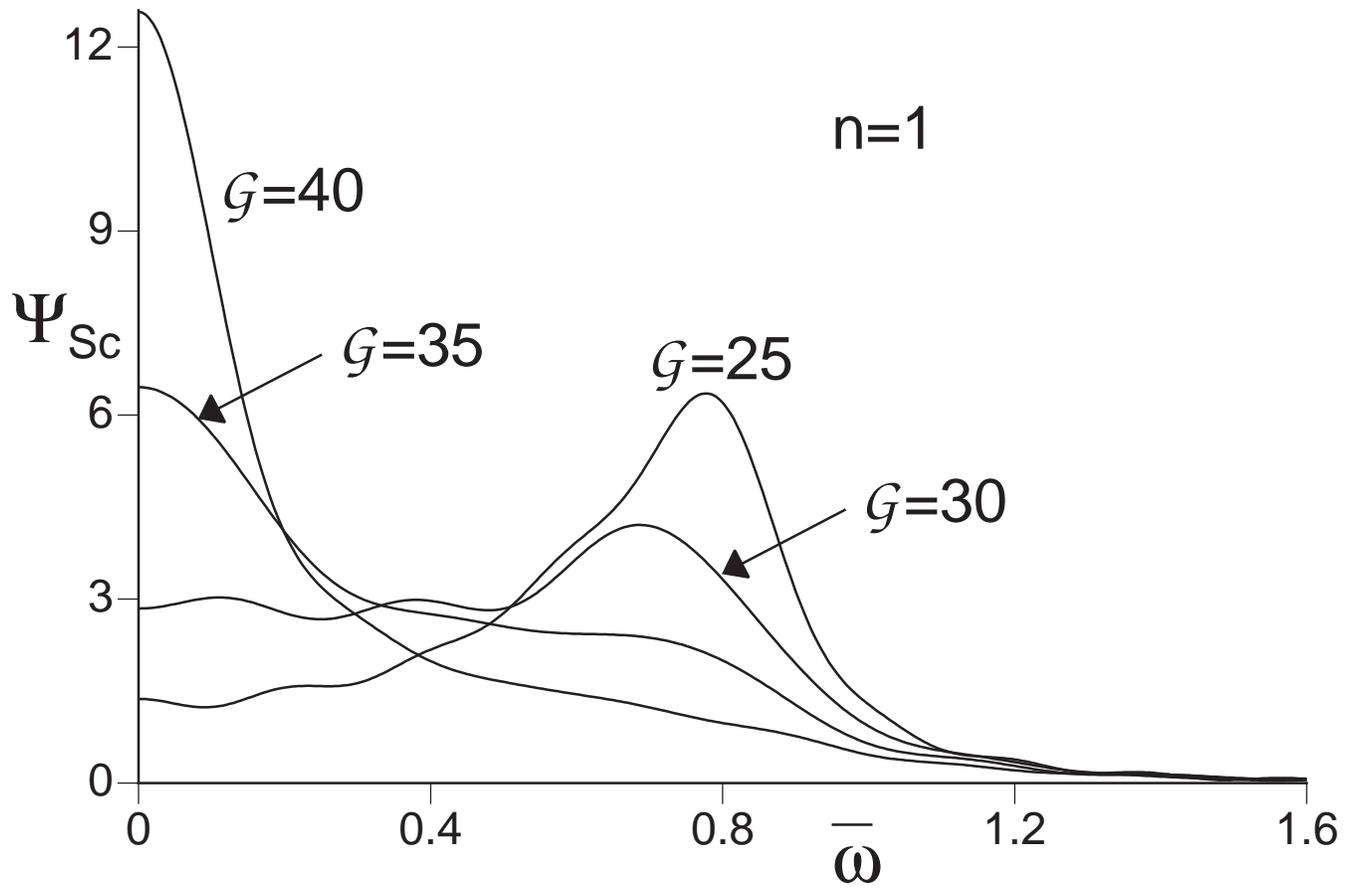}}
\caption{
The Fourier transform of Fig~\protect\ref{fig7} to frequencies,
which yields the scaling function $\Psi_{Sc} (0, \overline{\omega}, {\cal G})$
in (\protect\ref{r9}), where $\overline{\omega} = \omega/c
\protect\sqrt{R}$. Results are for $n=1$ and ${\cal G} = 25, 30, 35, 40$.
}
\label{fig10}
\end{figure}

\begin{figure}
\epsfxsize=7in
\centerline{\epsffile{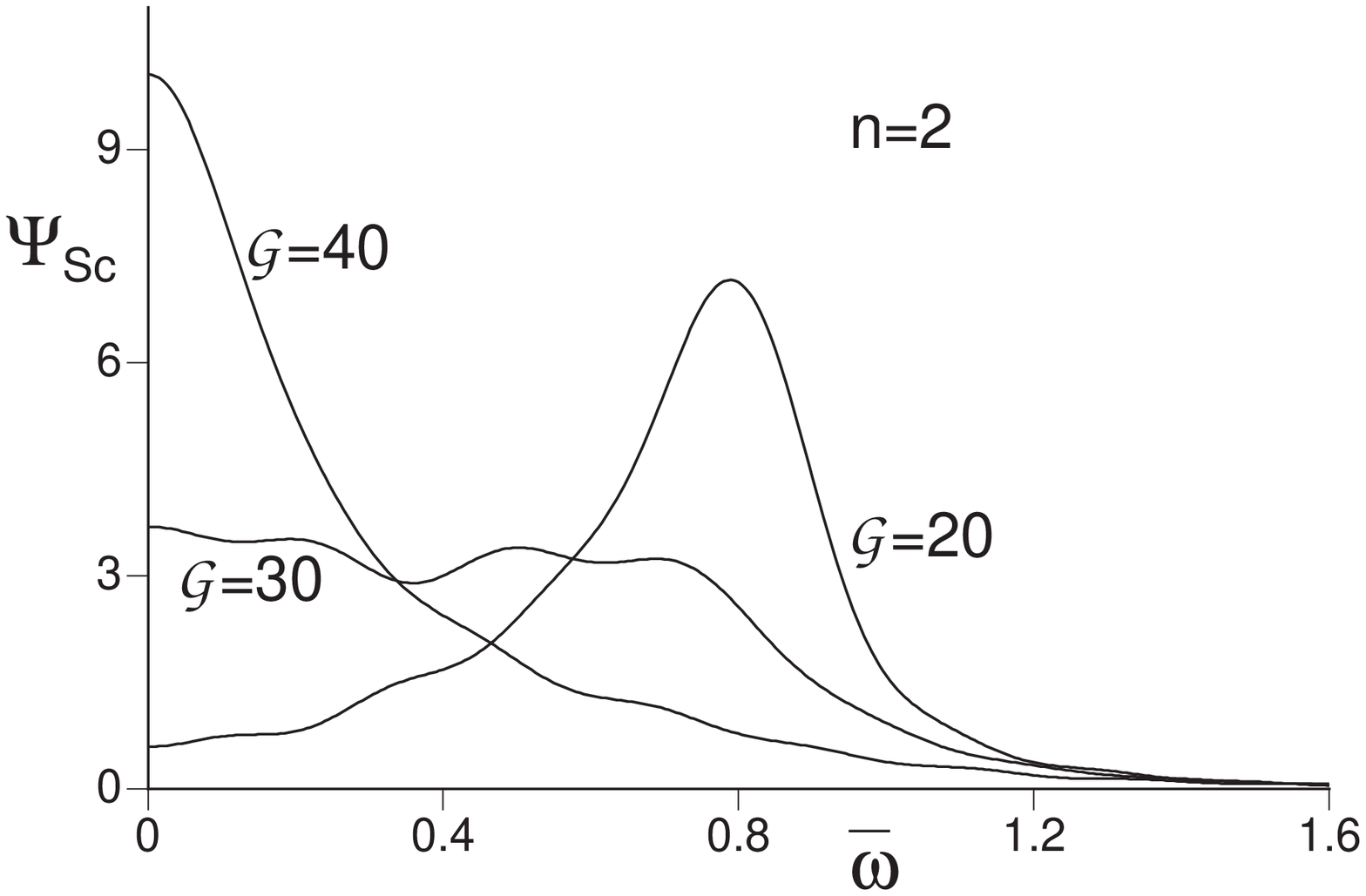}}
\caption{
As in Fig~\protect\ref{fig10} but for $n=2$ and ${\cal G} = 20, 30, 40$.
}
\label{fig11}
\end{figure}

\begin{figure}
\epsfxsize=7in
\centerline{\epsffile{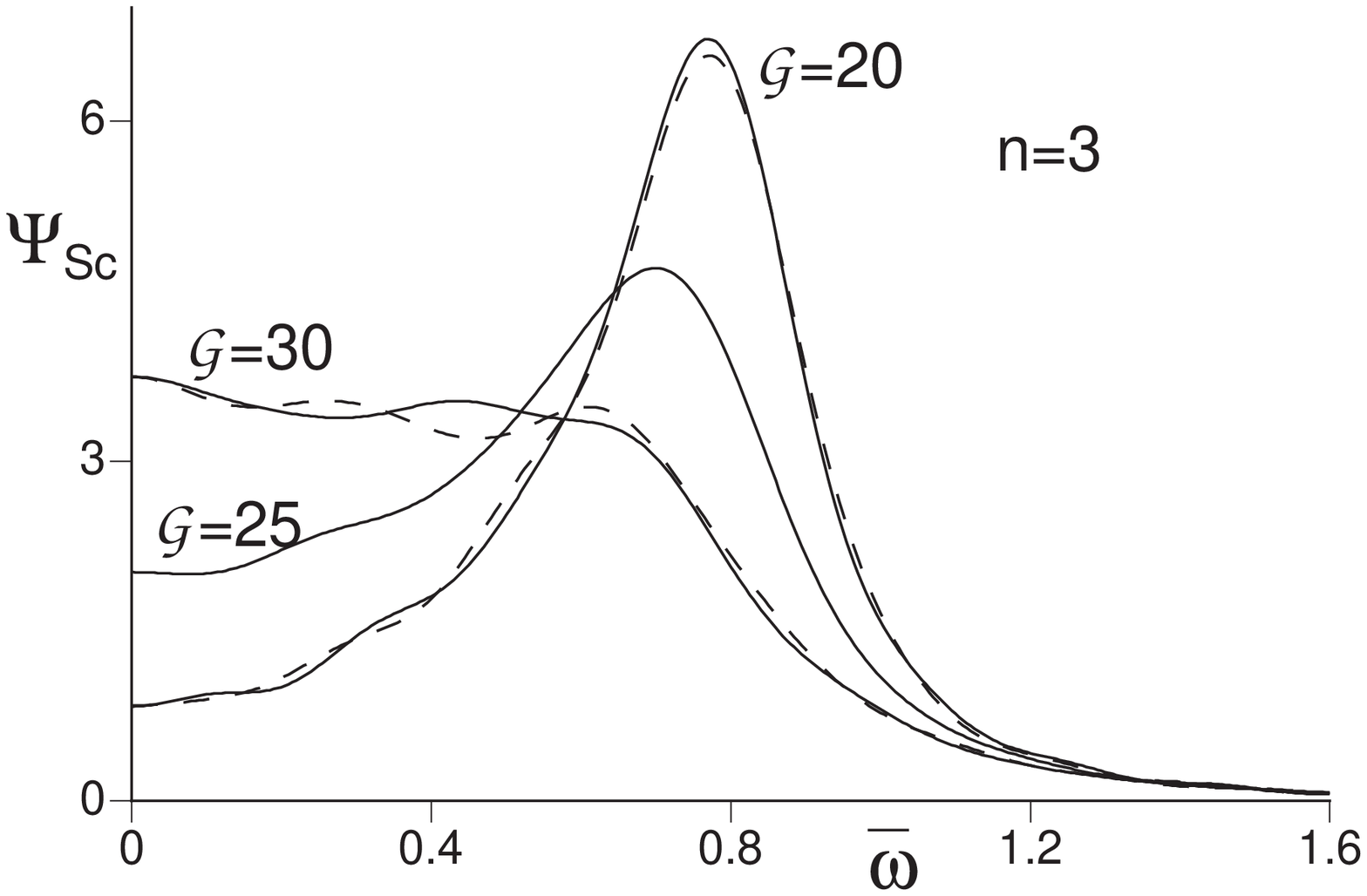}}
\caption{
As in Fig~\protect\ref{fig10} but for $n=3$ and ${\cal G} = 20, 25, 30$.
As in Fig~\protect\ref{fig9}, there are two data sets for ${\cal G}=20,30$
(indicated by the dashed and full lines),
corresponding to two different values of $Ra^2$.
}
\label{fig12}
\end{figure}

The Fourier transform of the data in Figs~\ref{fig7}--\ref{fig9}
to frequency directly gives us the dynamic
structure factor, and the scaling function $\Psi_{Sc}$ defined in
(\ref{r9}). The results for this are shown in Figs~\ref{fig10}--\ref{fig12}
for $n=1, 2$ and $3$ respectively.
Notice that $S(0,0)$ is clearly always non-zero. Consequently,
by the fluctuation-dissipation theorem, (\ref{fdt}) or
(\ref{cfdt}), $\mbox{Im} \chi (0, \omega) \sim \omega$ for small
$\omega$. The perturbative computations in I did not obey this
simple and important low frequency limit, and so this sickness has
been cured by the present non-perturbative, but numerical,
computation.

The small ${\cal G}$ regime of amplitude fluctuations discussed
above in the time domain, translates now into a peak in $S(0,\omega)$ at finite
frequency of order $\sim c \sqrt{R}$.
As ${\cal G}$ is reduced, we move out of the high $T$ region A in
Fig~\ref{fig1}, and into low $T$ region C on the quantum
paramagnetic side. This finite frequency, amplitude fluctuation
peak connects smoothly with a sharp peak associated with a quasiparticle
excitation of the quantum paramagnet.
Of course, once we are region C,
the amplitude and width of the peak can no longer be computed
by the present quasi-classical {\em wave} description,
and we need an approach which treats the excited {\em particles}
quasi-classically.

Now consider the opposite trend of increasing ${\cal G}$ towards
the low $T$ region B on the magnetically ordered side.
As ${\cal G}$ increases, the
peak broadens and eventually, the maximum moves down to zero
frequency. For $n \geq 2$, it is natural to interpret this dominance of low
frequency relaxation as due to ``phase'' or ``angular''
fluctuations of $\Phi_{\alpha}$ along the contour of zero energy
deformations at a fixed non-zero $|\Phi_{\alpha}|$. Of course the
fully renormalized effective potential necessary has a minimum at $\Phi_{\alpha} = 0$
because this is a region without long range order; nevertheless,
there must be a significant intermediate length scale over which
the local effective potential has a minimum at a non-zero value of
$|\Phi_{\alpha}|$, and the predominant fluctuations of $\Phi_{\alpha}$
are angular. This can also be seen from the relation
(\ref{defrt}): crudely, we can imagine varying $a$ at fixed $R$ to
determine the effective mass $\widetilde{R}$ on a length scale
$a$---we see that for large ${\cal G}$, there is a significant
scale over which $\widetilde{R}$ is negative, which prefers a
locally non-zero value of $|\phi_{\alpha}|$, and allows for
low-energy phase fluctuations.

We can understand the nature of the dynamics in the limit ${\cal G} \rightarrow \infty$
by arguments analogous to those made for the statics. We will restrict
our attention here to $n=3$, and other cases are similar. We saw in
Section~\ref{heisenberg} that the statics where described by the $d=2$
${\rm O}(3)$ non-linear $\sigma$-model. In a similar manner we can
argue that the dynamics will be given by the dynamical extension
of this model considered by Tyc {\em et al.}~\cite{tyc}, and the
three-argument universal scaling function $\Psi_{Sc}$
in (\ref{r9}) will collapse
to the two-argument universal scaling functions of Tyc {\em et
al.} in the limit ${\cal G} \rightarrow \infty$.
This collapse is similar to the transformation of (\ref{defpsi})
to (\ref{s11a}) in the same limit for the
statics. Consistent with the interpretation in the previous
paragraph, the description of the
${\cal G} \rightarrow \infty$ limit of the dynamics given by the
model of Tyc {\em et al} is described by a model in which
$|\phi_{\alpha}|$ is constrained to have a fixed length. Further
the results of Tyc {\em et al} show a large $\omega =0$ peak in
$S(0,\omega)$~\cite{tyc}, which is consistent with the trends
observed here with increasing ${\cal G}$.

The above description of the origin of the low frequency
relaxation in the continuum high $T$ region A due to angular
fluctuations in $\phi_{\alpha}$ clearly relies on the existence of
a continuous symmetry for $n \geq 2$. However, closely related
arguments can also be made for $n=1$ by appealing to the
low-energy mode arising from the motion of domain walls between
ordered regions with opposite orientations.

We have implicitly assumed above that `amplitude' and `angular'
fluctuations are mutually exclusive phenomena, but this is clearly
not true in principle. Even in a region with angular fluctuations,
there can be an amplitude mode involving fluctuations in $|\phi_{\alpha}|$
about its local potential minimum. Such a situation would be manifested by a
simultaneous peak in $S(0, \omega)$ both at $\omega =0$ and at a
finite frequency. It is apparent from Figs~\ref{fig10}-\ref{fig12}
that such a situation never arises in a clear-cut manner.
Apparently, once angular fluctuations
appear, the non-linear
couplings between the modes is strong enough in $d=2$ to reduce
the spectral weight in the amplitude mode to a small amount.
However the amplitude mode does not completely disappear---there
is a clearly visible shoulder in the $n=1$ Fig~\ref{fig10} for ${\cal G} = 35$,
indicating concomitant angular and amplitude fluctuations.
These results on the difficulty of observing an amplitude mode
for large ${\cal G}$ in $d=2$ (low $T$ in region B)
connect smoothly with the $T=0$ response of the magnetically ordered state of
the quantum
theory ${\cal Z}_Q$---the latter is reviewed in Appendix~\ref{ampq}, and
we find there that the amplitude mode is swallowed up in
the spin-wave continuum for the continuous symmetry case.

For the quantum-critical region A in Fig~\ref{fig1}, we should use
the value of ${\cal G}$ in (\ref{rest}). At $\epsilon=1$, this
result evaluates to ${\cal G} = 35.5$ for $n=2$,
${\cal G} = 29.2$ for $n=2$, and to ${\cal G}=24.9$
for $n=3$. If we take this value of ${\cal G}$ seriously, then we
see from Figs~\ref{fig10}--\ref{fig12} that all cases are quite
close to the border between amplitude and phase fluctuations,
when the peak in $S(0,\omega)$ moves from non-zero to zero frequency.
Amplitude fluctuations are however somewhat stronger for $n=3$
(when there is a well-defined peak at a non-zero frequency),
while angular/domain-wall relaxational dynamics is stronger
for $n=1$ (when there is a prominent peak at $\omega =0$).

There is a passing
resemblance between
the above crossover in dynamical properties as a function of ${\cal
G}$,
and a well-studied phenomenon in dissipative
quantum mechanics \cite{leggett,dissqm,lesage}: the crossover
from `coherent oscillation' to
`incoherent relaxation' in a two-level system
coupled to a heat bath . However, here we do not rely on an
arbitrary heat bath of linear oscillators, and the relaxational
dynamics emerges on its own from the underlying
Hamiltonian dynamics of an interacting many-body, quantum system.
Our description of the crossover has been carried out in
the context of a quasi-classical model, but, as we noted earlier,
the `coherent' peak
connects smoothly to the quasiparticle peak in region C of
Fig~\ref{fig1}; in this latter region the wave oscillations get
quantized into discrete lumps which must then be described by
a `dual' quasi-classical particle picture.

\section{Implications for experiments}
\label{expts}

We first summarize the main theoretical results of this paper.
It is convenient to do this in two steps: first for the statics,
and then the dynamics.

For static properties, we presented a rather complete analysis of the
classical field theory, ${\cal Z}$ in (\ref{cal}), of a $n$ component
scalar field $\Phi_{\alpha}$ with a quartic self interaction.
We motivated the study of ${\cal Z}$ here as an effective theory of
the static fluctuations of the quantum model ${\cal Z}_Q$, in (\ref{calsq}),
but it is clear that ${\cal Z}$ has a much wider domain of applicability.
The static properties of almost any quantum model
in $d=2$ with an ${\rm O}(n)$-symmetric order
parameter should be satisfactorily modeled by ${\cal Z}$. Of course, the
$T$-dependent values
of the coupling constants, $R$ and ${\cal G}$ will then be different, and depend
upon the specific underlying quantum model. So, for instance, if one
of the phases was a $d$-wave superconductor, and had gapless fermionic
excitations, then the subleading corrections to
expressions for the couplings $R$ and ${\cal G}$ in (\ref{app16})
would change. Apart from a traditional weak coupling analysis of ${\cal
Z}$, we introduced a surprising, exact solution of the strong
coupling limit by a duality-like transformation. We also
interpolated between these limits by Monte Carlo simulations.
Among our main new results was a computation of the $T$ dependence
of the spin stiffness, $\rho_s (T)$ for $n=2$, and shown in
Fig~\ref{fig4}. When combined a knowledge of the temperature
dependence of ${\cal G}$ as in the scaling form (\ref{phig}) it
leads to a prediction for $\rho_s (T)$ consistent with the form (\ref{r1}).
As a first pass, we can combine the approximate low $T$ prediction
which follows from (\ref{app16}), ${\cal G}_D \approx 3T/\rho_s
(0)$, with Fig~\ref{fig4} and obtain an explicit prediction for
the function $\Psi_{\rho}$ in (\ref{r1}).

Our studies of dynamic properties were somewhat more specialized
to the quantum model ${\cal Z}_Q$, although we expect
that closely related methods can be applied to other models.
Our approach relied heavily on the specific value of the
`mass' $R$ obtained in the high $T$ limit of ${\cal Z}_Q$---we
used the fact that $c \sqrt{R}/T \sim \sqrt{\epsilon} \ll 1$ to argue for
an existence of a dynamical model of classical waves. This model
is defined by the ensemble of initial conditions (\ref{calc}) and
the equations of motion (\ref{r6}) and (\ref{r7}). The universal dynamical
properties of this model were studied by numerical simulations
directly in $d=2$, and results are summarized in
Figs~\ref{fig7}--\ref{fig12}. As we pass from region C to A to B
in Fig~\ref{fig1}, the value of the coupling ${\cal G}$ increases
monotonically. The dynamics shows a crossover from
a finite frequency, ``amplitude fluctuation'', gapped quasiparticle mode
to a zero frequency ``phase'' ($n \geq 2$) or ``domain wall'' ($n=1$)
relaxation mode during this increase in ${\cal G}$.

One of the primary applications of our results is to the
superfluid-insulator transition for the case $n=2$. Our approach offers a
precise and well-defined method to describe the
physics of strongly fluctuating superfluids above their critical
temperature, with an appreciable density of vortices present.
Near a quantum critical point, our results show that
such superfluids (region A of Fig~\ref{fig1}) have a
reasonably well-defined order parameter, with $|\Phi_{\alpha}|$
non-zero over a significant intermediate length scale. Strong
phase fluctuations~\cite{ek} are eventually responsible for the
disappearance of true long-range order. The evidence for this
picture comes from our dynamic simulations, which show a
well-formed relaxational peak at {\em zero} frequency in
the dynamic structure factor. The trends in the dynamic structure
factor as a function of ${\cal G}$ then support the interpretation
that this peak arises from phase relaxation.
We speculate that
our results can be extended to deduce consequences for the
electron photo-emission spectrum of the high temperature
superconductors, along the lines of Refs~\cite{franz,dorsey}. We
imagine, in a Born-Oppenheimer picture, that the fermionic quasiparticles are moving in a
quasi-static background of the $\Phi_{\alpha}$ field. Then the
photoemission cross section can be related to a suitable
convolution of the electron Green's function and the dynamic
structure factor of $\Phi_{\alpha}$. Under such circumstances, we
believe that a zero frequency, phase relaxation peak in the
dynamic structure factor of $\Phi_{\alpha}$ will translate into a
weak `pseudo-gap' in the fermion spectrum.

A separate application of our results is to zero temperature
magnetic disordering transitions for the case $n=3$.
Recent neutron scattering measurements of
Aeppli {\em et al.}~\cite{aeppliscience} on ${\rm La}_{2-x}
{\rm Sr}_x {\rm Cu O}_4$ at $x=0.15$ are consistent with
quantum-critical scaling with dynamic critical exponent
$z = 1$ and anomalous field exponent $\eta \approx 0$,
suggesting proximity to an insulating state with
incommensurate, collinear
spin and charge ordering (the collinear spin-ordering ensures that
a single $n=3$ vector order parameter is adequate; coplanar ordering would require
a more complex order parameter and is not expected to have $\eta$
close to 0).
Their measurements have so far mainly focused on the momentum
dependence of the structure factor, and are well fit by a
Lorentzian squared form. We have not computed such momentum
dependence in our simulations here, but general experience with
scaling functions in $d=1$ suggests that such a form is to be
expected in the high $T$ regime A~\cite{seoul}.
It would be quite interesting to examine the $\omega$ dependence
of the structure factor in future experiments, and compare them
with our results in Fig~\ref{fig12}.
As we discussed in Section~\ref{dyn},
it is the smaller values of ${\cal G}$, which have
a {\em non-zero} frequency peak in $S(0,\omega)$ in Fig~\ref{fig12},
and which lead to a
`pseudo-gap' in the {\em spin} excitation spectrum.
Compare this with our discussion of the $n=2$ superfluid-insulator
transition above, where a {\em zero} frequency peak in
$S(0, \omega)$ at larger values of ${\cal G}$
was argued to lead to a {\em fermion} pseudo-gap.
It is satisfying to note that if we are to observe both a spin and a
fermion `pseudo-gap' in the experiments, then the trend in the required values of
${\cal G}$ with $n$ is consistent with (\ref{rest}).

Continuing our discussion of the application of $n=3$ to the high
temperature superconductors, it is also appears worthwhile to
remind the reader of nature of the magnetic spectrum in the
magnetically disordered side, in region C of Fig~\ref{fig1}~\cite{RS,CSY}.
Here
there is a sharp triplet particle excitation above the spin gap,
which is only weakly damped. It is interesting that such
excitations have been observed at low $T$ in ${\rm Y Ba}_2 {\rm Cu
O}_{6+x}$~\cite{keimer}. However, their eventual interpretation must
await more detailed studies and comparison with the situation in
${\rm La}_{2-x}
{\rm Sr}_x {\rm Cu O}_4$.

It is tempting to combine our discussion of $n=2$ and $n=3$ models
above for the high temperature superconductors, into a single $n=5$
model~\cite{zhang}. However, there does not appear to be any reason for the resulting
theory to be even approximately ${\rm O}(5)$ invariant. Moreover,
we have seen above that we need the freedom to independently vary ${\cal G}$
for the $n=2$ and $n=3$ subsystems, at moderately large values,
to obtain a proper description
of the physics; for $n=5$, the value of ${\cal G}$ in (\ref{rest})
is quite small, and would lead to a rather sharp gap-like
structure in the dynamic structure factor of the $n=5$ order parameter
in the high $T$ limit.

Finally, we mention recent light scattering experiments~\cite{pelle} on
double layer quantum Hall systems which have explored both sides
of a magnetic ordering transition. Simulations at $n=3$, but in
the presence of a magnetic field can lead to specific predictions
for this system. The experimental results appear to show the appearance of a
spin pseudo-gap at high $T$, which is consistent with our
results so for $n=3$.

\acknowledgements
I thank G.~Aeppli, C.~Buragohain, K.~Damle, J.~Tranquada and J.~Zaanen for
useful discussions.
This research was supported by NSF Grant No DMR 96--23181.

\appendix

\section{Spectrum of the ordered state of the quantum theory at $T=0$}
\label{ampq}

We address here the issue of amplitude fluctuations of $|\phi_{\alpha}|$
in a state with magnetic long-range order. We will do this by
examining the response functions of the quantum theory ${\cal Z}_Q$
in (\ref{calsq}) at $T=0$. These were computed in I by the $\epsilon$
expansion---for small $\epsilon$, there is a well-defined peak in
the spectral density of the longitudinal response functions,
corresponding to amplitude oscillations of $|\phi_{\alpha}|$
about a non-zero value. Indeed, these results were used by Normand
and Rice~\cite{normand} to argue that such an amplitude mode will
be observable in the insulator ${\rm La Cu O}_{2.5}$, which is
believed to be near a $d=3$ quantum critical point.

We will consider the case $d=2$ here, using the large $n$
expansion. Unlike $d=3$, we will find here that such
an amplitude mode is not visible in $d=2$ because the
cross-section for decay into multiple spin-wave excitations is too
large. This result also connects smoothly with our $T>0$, $d=2$
simulations in Section~\ref{dyn}, where again we found little sign
of such an amplitude mode.

Large $n$ results for the two-point $\phi_{\alpha}$ correlator in
the direction parallel ($\chi_{\parallel} (k, \omega)$) and
orthogonal ($\chi_{\perp} (k, \omega)$) to the spontaneous
magnetization were given in Appendix D of Ref~\cite{CSY}.
They are {\em universal} functions of $\rho_s (0)$, $c$,
and the ground state spontaneous magnetization, $N_0$, and in
particular, they do not depend upon whether the underlying degrees
of freedom are soft spins (as in ${\cal Z}_Q$) or vectors of unit
length (as in Ref~\cite{CSY}).
To leading order in $1/n$, the results are
\begin{eqnarray}
\chi_{\perp} (k, \omega) &=& \frac{N_0^2}{\rho_s (0) \left[k^2 -
(\omega/c)^2 \right]} \nonumber \\
\chi_{\parallel} (k, \omega) &=& \frac{N_0^2}{\rho_s (0)}
\frac{1}{\sqrt{k^2 - (\omega/c)^2} \left[ \sqrt{k^2 - (\omega/c)^2} +
16 \rho_s (0)/ c n\right]}.
\label{ampq1}
\end{eqnarray}
So, as expected, the transverse correlator has a simple pole at the spin-wave
frequency, $\omega = c k$. On the other hand, the longitudinal
correlator only has a branch-cut at $\omega = c k$---the spectral
density vanishes for $\omega < c k$, and decreases monotonically for $\omega > c k$. In
particular, there is no pole-like structure at a frequency of
order $\rho_s (0) /n$, which is the expected position of the amplitude
mode.

\section{Computation of $R$ and $U$}
\label{ru}
This appendix discuss the values of $R$ and $U$ obtained by
integrating out the non-zero imaginary frequency modes from ${\cal
Z}_Q$. Such a computations was already discussed in I, but here we
will present the modifications necessary due to the slightly
different renormalization used in (\ref{defR}). We will also
present new computations within the magnetically ordered state in
region B of Fig~\ref{fig1}.

First, in Section~\ref{ru1}, we will follow the paramagnetic
approach of I, which computes
parameters for $r>0$, and then extrapolates to $r<0$ by a method of
analytic continuation in $r$, which is valid for $T>0$. This
method works without any hitches in regions A and C of
Fig~\ref{fig1}. In principle, it is also expected to be valid within all
of region B, but the results becomes progressively poorer as the
limits $T \rightarrow 0$ and $\epsilon \rightarrow 0$ do not
commute for $r<0$.
We will then present, Section~\ref{ru2} an alternate computation which begins within
the magnetically ordered state of region B and then directly
computes the `dual' couplings $R_D$ and $U$.

\subsection{Paramagnetic approach}
\label{ru1}

We begin by noting that the value of the bare coupling $r_c$
in (\ref{calsq}) is
\begin{equation}
r_c = - u \left(\frac{n+2}{6} \right) \int \frac{d^d k}{(2\pi)^d}
\int \frac{d \omega}{(2 \pi)} \frac{1}{k^2 + (\omega/c)^2},
\label{app1}
\end{equation}
to leading order in $u$.
We assume that $r>0$, and so it is valid to integrate out
fluctuations in $\phi_{\alpha}$ about $\phi_{\alpha} = 0$.
Integrating out the $\omega_n \neq 0$
modes from ${\cal Z}_Q$ in this manner to leading order in $u$, and comparing
resulting effective action with ${\cal Z}$, we find
straightforwardly
\begin{eqnarray}
\widetilde{R} &=& r + u \left( \frac{n+2}{6} \right) \int
\frac{d^d k}{(2 \pi)^d} \left[ T \sum_{\omega_n \neq 0}
\frac{1}{k^2 + (\omega_n /c)^2 + r} - \int \frac{d \omega}{(
2\pi)}
\frac{1}{k^2 + (\omega/c)^2} \right] \nonumber \\
U &=& u - u^2 \left( \frac{n+8}{6} \right) \int
\frac{d^d k}{(2 \pi)^d} T \sum_{\omega_n \neq 0}
\frac{1}{(k^2 + (\omega_n /c)^2 + r)^2}
\label{app2}
\end{eqnarray}
The remaining task is, in principle, straightforward: we have to combine
(\ref{app2}) with (\ref{defR}), and evaluate the resulting
expressions to obtain our final results for $R$ and $U$.
In the vicinity of the quantum critical point in Fig~\ref{fig1},
the resulting expressions should be universal functions only of
$T$, $c$, and an energy scale measuring the deviation of the
ground state couplings from the $r=0$ point; for this energy scale we
choose either the ground state spin stiffness, $\rho_s (0)$ (for $r<0$
and $n > 1$), or the ground state energy gap (which is
$\Delta_-$ for $r<0$, $n=1$, and is $\Delta_+$ for $r>0$ and all
$n$).

One reasonable approach at this point is to solve
(\ref{app2}) and (\ref{defR})
for $R$ and $U$ directly in $d=2$ by a numerical method.
This will give results for $R$ and ${\cal G}$ which
are valid everywhere in the phase diagram of Fig~\ref{fig1}---the
resulting value of ${\cal G}$ will increase monotonically from 0
to $\infty$ as one moves from region C to A to B. Moreover, $R$
will remain positive everywhere. However, the results will not be
explicitly universal and will depend upon microscopic parameters
from the theory ${\cal Z}_Q$---the values of $u$ and the momentum
cutoff $\Lambda$.

Universality of the final result can only be established order by order
in $\epsilon$. In the remainder of this subsection we will
evaluate (\ref{app2}) in such an
expansion in $\epsilon$. This surely reduces the accuracy of our
final estimates for $R$ and $U$; it should not be forgotten,
however, that we subsequently study ${\cal Z}$ directly in $d=2$,
and so the scaling functions in (\ref{defpsi}) and (\ref{r9}),
whose arguments are related to $R$
and $U$, are known much more accurately.
We will find that our leading order in $\epsilon$ result for
$R$ vanishes when $T$ becomes sufficiently small in region B---then
the $\epsilon$ expansion can no longer be considered adequate for
estimating $R$ in $d=2$. The vanishing of $R$ is acceptable for
small $\epsilon$ ($d>2$) however, because we can then do a massless subtraction in
(\ref{defR}) and negative values of $R$ merely place the system
well within the magnetically ordered state. An alternative,
$\epsilon$-expansion
computation of $R$ for systems well within region B will appear in
the following subsection.

The techniques for reducing (\ref{app2}) and (\ref{defR}) into a
universal form in the $\epsilon$ expansion have been discussed at
length in I, and also in Ref~\onlinecite{DSZ}.
Using these methods we find to leading order in $\epsilon$
\begin{eqnarray}
&& c^2 R + 2 \pi \epsilon \left( \frac{n+2}{n+8} \right) c T \sqrt{R}
= c^2 r \left[ 1 + \epsilon \left( \frac{n+2}{n+8} \right)
\ln \left(\frac{T}{c\mu}\right) \right]+
\epsilon T^2 \left( \frac{n+2}{n+8} \right) G \left(
\frac{c^2 r}{T^{1/z\nu}} \right)  \nonumber \\
~~~~&& c U =
\frac{6 \epsilon (T/c)^{(3-d)}}{(n+8) S_{d+1}} \left[
1 + \epsilon \frac{(20+2n-n^2)}{2(n+8)^2} + \epsilon G'\left(
\frac{c^2 r}{T^{1/z\nu}} \right) \right],
\label{app3}
\end{eqnarray}
where $S_d = 2/[\Gamma(d/2) (4 \pi)^{d/2}]$ a phase space factor,
$\nu = 1/2 + \epsilon (n+2)/(4 (n+8))$ to this order in the $\epsilon$
expansion,
and $\mu$ is a short-distance momentum scale which can be
eliminated by re-expressing $r$ in terms of physical energy
scales. The function $G(y)$ was given in (D8) of II:
\begin{equation}
G(y) = - 2 \int_0^{\infty} d q \left[
\ln \left( 2 q^2 \frac{(\cosh (\sqrt{q^2 + y}) - 1)}{q^2 + y} \right)
-q - \frac{y}{2 \sqrt{q^2 + 1/e}} \right]
\label{app3a}
\end{equation}
This form of $G(y)$ is valid for both negative and positive $y$ (when the argument
of the square root is negative we use the identity $\cosh ix = \cos x$)
and is easily shown to be analytic at $y=0$ where
\begin{equation}
G(0) = \frac{2 \pi^2}{3}~~~~,~~~~dG/dy (0) = 2.453808582\ldots .
\label{app3b}
\end{equation}
The simple form $G(y) \approx G(0) + y (dG/dy(0))$ is actually a
reasonable approximation for $G(y)$ over a wide range of values of
$y$. For $y \rightarrow \infty$, we can show from (\ref{app3a})
that
\begin{equation}
G(y \rightarrow \infty) = \frac{y \ln y}{2} + 2 \pi \sqrt{y} +
\sqrt{8 \pi \sqrt{y}} e^{-\sqrt{y}} + \ldots
\label{app3c}
\end{equation}

The expression (\ref{app3}) for $U$ is identical to that in I,
while that for $R$ differs only in that the term proportional to
$T \sqrt{R}$ on the left hand side
was absent in I. This difference is of course due to the
subtraction term with mass $R$ in (\ref{defR}). Because of the
presence of this term, (\ref{app3}) is reasonable only as long
as there is a solution with $R>0$. For low enough $T$ in region B
there will be no such solution, and then the present method
breaks down as a method for estimating $R$ in $d=2$; an
alternative approach will be discussed in Section~\ref{ru2}.

We complete this appendix by relating $r$ to physical energy
scales to leading order in $\epsilon$. The reader can easily
verify that when we eliminate $r$ in (\ref{app3}) by
the following expressions, the arbitrary scale $\mu$ disappears to
the appropriate order in $\epsilon$. The following relations can
also be used in to similarly eliminate $r$ from the expressions in
Section~\ref{ru2}.

For $r>0$, the ground state has an energy gap, $\Delta_+$ for all
$n$. Then, from I, we have
\begin{equation}
\Delta_+ = c\mu \left(\frac{r}{\mu^2}\right)^{\nu}.
\label{app4}
\end{equation}

For $r<0$, we have to distinguish $n=1$ and $n \geq 2$. For $n=1$,
there is an energy gap, $\Delta_-$, and
\begin{equation}
\Delta_- = c\mu \left( 1 + \frac{\pi \sqrt{3} - 3}{12} \epsilon
\right) \left( \frac{-2 r}{\mu^2} \right)^{\nu}.
\label{app5}
\end{equation}
For $n \geq 2$ it is convenient to use the parameter
$\widetilde{\rho}_s$ obtained from the ground state spin
stiffness, which has the engineering dimensions of energy in all
$d$ (in $d=2$ it is simply proportional to $\rho_s (0)$):
\begin{equation}
\widetilde{\rho}_s = c^{(d-2)/(d-1)} \left( \frac{2 \epsilon}{(n+8)}
\frac{\rho_s (0)}{S_{d+1}} \right)^{1/(d-1)}.
\label{app6}
\end{equation}
Then
\begin{equation}
\widetilde{\rho}_s = c \mu \left(1 - \epsilon \frac{12+n-2n^2}{4 (n+8)^2}
\right) \left( \frac{-2 r}{\mu^2} \right)^{\nu}.
\label{app7}
\end{equation}

\subsection{Magnetically ordered approach}
\label{ru2}

Here we will directly compute the `dual' couplings $R_D$ and $U$
by matching the effective potential for $\Phi_{\alpha}$
fluctuations in ${\cal Z}$ in (\ref{cal})
with that obtained by integrating out
the nonzero Matsubara frequency modes in ${\cal Z}_Q$ in (\ref{calsq}).
We will initially assume that we are
in the magnetically ordered state, and so $r<0$ and $\widetilde{R}<0$.
The effective potential
of (\ref{cal}) has a
minimum at $|\Phi_{\alpha}|^2 = - 6 \widetilde{R}/U$, and its
curvature at this minimum is $-2 \widetilde{R}/T$. We compute
precisely the same quantities from the free energy obtained from ${\cal Z}_Q$
after integrating out the nonzero Matsubara frequency modes at the
one loop level, and so obtain expressions for $\widetilde{R}/U$ and
$\widetilde{R}/T$. We solve these for $\widetilde{R}$ and $U$, and
obtain results which replace (\ref{app2}):
\begin{eqnarray}
&& \widetilde{R} = r + u \int \frac{d^d k}{(2 \pi)^d} T
\sum_{\omega_n \neq 0} \left[ \frac{k^2 + (\omega_n/c)^2 - 5r}{2 (k^2 + (\omega_n/c)^2
-2 r)^2} + \frac{(n-1)(k^2 + (\omega_n/c)^2 - r)}{6 (k^2 +
(\omega_n/c)^2)^2} \right] \nonumber \\
&&~~~~~~~~~~~~~~~~~~~~~- u \left( \frac{n+2}{6} \right) \int \frac{d^d k}{(2 \pi)^d}
\int \frac{d \omega}{2 \pi} \frac{1}{k^2 + ( \omega/c)^2}
\nonumber \\
&& U = u - u^2 \int \frac{d^d k}{(2 \pi)^d} T
\sum_{\omega_n \neq 0} \left[ \frac{3}{(k^2 + (\omega_n/c)^2
-2 r)^2} + \frac{n-1}{6 ( k^2 + (\omega_n/c)^2 )^2} \right]
\label{app8}
\end{eqnarray}
We now use (\ref{defQ}) and evaluate these expressions by the same
methods which led to (\ref{app3}). This replaces (\ref{app3}) by
\begin{eqnarray}
&& \frac{c^2 R_D}{2} - 2 \pi \epsilon \left( \frac{n+2}{n+8} \right) c T \sqrt{R_D}
= -c^2 r \left[ 1 + \epsilon \left( \frac{n+2}{n+8} \right)
\ln \left(\frac{T}{c\mu}\right) \right]-
\frac{\epsilon T^2}{n+8} G_1 \left(
-\frac{c^2 r}{T^{1/z\nu}} \right) \nonumber \\
~~~~&& c U =
\frac{6 \epsilon (T/c)^{(3-d)}}{(n+8) S_{d+1}} \left[
1 + \epsilon \frac{(20+2n-n^2)}{2(n+8)^2}
 + \frac{\epsilon}{(n+8)} G_2\left(
-\frac{c^2 r}{T^{1/z\nu}} \right) \right],
\label{app9}
\end{eqnarray}
where
\begin{eqnarray}
G_1 (y) &=& 3 G(2y) + (n-1)G(0) - 9yG'(2y) - (n-1) y G'(0) \nonumber
\\
G_2 (y) &=& 9 G'(2y) + (n-1) G'(0).
\label{app10}
\end{eqnarray}
There expressions can now be combined with (\ref{app5}) and
(\ref{app7}) to obtain universal expressions for $R_D$ and $U$
within region B, and also into the crossover into region A.
The resulting values, when combined with (\ref{s1}) will not agree
precisely with those obtained from (\ref{app3}) because the
computations in the present appendix are good to leading order in
$\epsilon$, while the relation (\ref{s1}) is only valid in $d=2$.

Let us look at the values of $R_D$ and $U$ obtained from
(\ref{app9}) in the limit $T \rightarrow 0$ with $r<0$. After using
(\ref{app3c}) it is straightforward to show that
\begin{eqnarray}
&& R_D = -2r \left[1 + \frac{9\epsilon}{2(n+8)}+
\epsilon \left( \frac{n-1}{n+8} \right)
G'(0) + \epsilon \left( \frac{n+2}{n+8} \right)
\ln \left(\frac{T}{c\mu}\right) + \frac{3 \epsilon}{2(n+8)}
\ln \left(\frac{-2c^2 r}{T^2} \right)\right] \nonumber \\
&&~~~~~~~~~~~~~~~~~~~+ \frac{\pi \epsilon T}{c} \left( \frac{4n+5}{n+8} \right) \sqrt{-2 r}
- \frac{2\epsilon T^2}{c^2} \left( \frac{n-1}{n+8} \right)
G(0) + \ldots \nonumber\\
&& c U =
\frac{6 \epsilon (-2r)^{(3-d)/2}}{(n+8) S_{d+1}} \left[
1 + \epsilon \frac{(20+2n-n^2)}{2(n+8)^2} \right. \nonumber \\
&&~~~~~~~~~~\left.+ \frac{\epsilon}{2(n+8)} \left\{
-(n-1) \ln \left(\frac{-2c^2 r}{T^2} \right)
+ 9 + 2(n-1)G'(0) + \frac{18\pi T}{c\sqrt{-2r}}
\right\} \right] + \ldots
\label{app11}
\end{eqnarray}
where only exponentially small terms of order $e^{-c\sqrt{-r}/T}$ have
been omitted. The $\ln (T)$ terms above are dangerous as they
do not have a finite limit as $T \rightarrow 0$. For $n=1$,
a glance at (\ref{app11}) shows that all such terms do indeed
cancel; so the result (\ref{app9}) remains valid everywhere in region B,
including in the limit $T \rightarrow 0$. Further it can be
checked (using results in I)
that ${\cal G}_D (T \rightarrow 0) = 3T/N_0^2$, which agrees with
(\ref{s7aa}).

However, for $n \geq 2$, there is no cancellation of such
terms. This indicates a breakdown of the $\epsilon$ expansion for
these values of $n$ as $T \rightarrow 0$.
The low-lying excitations for these cases are gapless
spin waves. In the present approach, by matching the quantum
theory to the classical action ${\cal Z}$, we are effectively
treating these spin waves as classical up to a high energy cutoff
of $c \sqrt{-2r}$, which is the mass of the amplitude mode. In
reality, however, such spin waves are only classical up to an
energy $k_B T$~\cite{CHN,CSY}, and this is responsible for the
breakdown of (\ref{app9}).

In the remainder of this subsection, we abandon the $\epsilon$
expansion, and use some reasonable physical criteria (suggested by the above
discussion) to estimate ${\cal G}_D$
and $R$ as $T \rightarrow 0$. Such estimates are, by nature,
unsystematic, and there does not appear to be any clear-cut procedure by
which they can be improved to extend to higher $T$.
We will match the exact large ${\cal G}$ form
of the correlators of ${\cal Z}$ in Sections~\ref{xy}
and~\ref{heisenberg} with known exact results for the low $T$
correlators of the quantum theory ${\cal Z}_Q$.
These quantum correlators can be universally expressed in terms of
$\rho_s (0)$, $c$, and the ground state spontaneous magnetization $N_0$

For $N=2$, the predominant fluctuations at low $T$ are angular
fluctuations about some locally ordered state. So we write
\begin{equation}
\Phi_1 + i \Phi_2 \sim e^{i \theta}.
\label{app12}
\end{equation}
The quantum action controlling fluctuations of $\theta$ is
\begin{equation}
{\cal L}_{\theta} = \frac{\rho_s (0)}{2} \left[ (\nabla_x
\theta)^2 + (\partial_{\tau} \theta)^2 \right].
\label{app13}
\end{equation}
Evaluating the two-point correlator from (\ref{app12}) and
(\ref{app13}), and fixing the missing normalization in (\ref{app12}) by matching
to the actual value of $N_0$, we obtain
\begin{equation}
\sum_{\alpha=1}^n \langle \Phi_{\alpha} (x) \Phi_{\alpha} (0) \rangle =
N_0^2 \exp \left[ \frac{1}{\rho_s (0)} \int \frac{d^2 k}{(2
\pi)^2} \left( T \sum_{\omega_n} \frac{e^{ikx} - 1}{k^2 +
(\omega_n/c)^2} + \int \frac{d \omega}{2 \pi} \frac{1}{k^2 +
(\omega/c)^2} \right)\right]
\label{app14}
\end{equation}
The expression (\ref{app14}) is free of both ultra-violet and
infra-red divergences, and we obtain for large $|x|$
\begin{equation}
\sum_{\alpha=1}^n \langle \Phi_{\alpha} (x) \Phi_{\alpha} (0) \rangle =
N_0^2 \exp \left[ - \frac{T}{2 \pi \rho_s (0)} \left( \ln(T |x|/c)
-\ln2 + \gamma \right) \right],
\label{app15}
\end{equation}
where $\gamma$ is Euler's constant.
Matching this correlator with that of the classical theory
${\cal Z}$ in (\ref{dual5}), we obtain two important, and exact
results
\begin{eqnarray}
{\cal G}_D (T \rightarrow 0) = \frac{3T}{\rho_s (0)} + \ldots
\nonumber \\
R_D (T \rightarrow 0) = \frac{T^2}{c^2} + \ldots
\label{app16}
\end{eqnarray}
where the ellipses represent unknown terms which are higher order in $T$.

A similar computation can be carried out for $n \geq 3$. In this
case, comparing the classical result (\ref{dual12}) with the
quantum results of Refs~\cite{CHN,CSY}, we find precisely the
relations (\ref{app16}). So (\ref{app16}) holds for all $n \geq
2$.

\section{Static correlations in the `dual' theory}
\label{dual}
This Appendix will show how we can compute correlators of the
theory (\ref{cal}) in the limit that the Ginzburg parameter
${\cal G}$ becomes very large. This is done in the `dual'
formulation discussed in Section~\ref{statstrong}: is a theory
characterized by a Ginzburg parameter ${\cal G}_D$ which becomes
small when ${\cal G}$ becomes large. All of the results will be
expressed in a perturbation theory in ${\cal G}_D$.

We first perform a naive computation of the ${\rm O}(n)$-invariant
two-point $\Phi_{\alpha}$
correlator in the original static theory (\ref{cal}), assuming
$\widetilde{R} < 0$. We will then renormalize it using
(\ref{defQ}), and will find, as expected, that the result is finite in the limit
$a \rightarrow 0$.
The interpretation of the result will be straightforward in the
for the case $n=1$ and will lead directly to (\ref{s6}).
We will then turn to further computations
for the cases $n=2$, and $n \geq 3$ in subsequent subsections.
An important ingredient in the interpretation of the results for
these cases shall be the computation of the change in the free
energy of the theory (\ref{cal}) in the presence of an external
field which couples to the generator of ${\rm O}(n)$ rotations: this
will allow us to compute the renormalized spin stiffness for
$n=2$, and the precise correlation length for $n \geq 3$.

At the mean-field level, for $\widetilde{R} < 0$ (which is assumed throughout
this appendix), $\Phi_{\alpha}$
fluctuates around the average value $|\Phi_{\alpha} | = \sqrt{6
|\widetilde{R}|/U}$. So we write
\begin{equation}
\Phi_{\alpha} (x)  = \sqrt{\frac{6 |\widetilde{R}|}{U}} \delta_{\alpha,1}
+ \widetilde{\Phi}_{\alpha} (x),
\label{dual1}
\end{equation}
where $\widetilde{\Phi}_{\alpha}$ represents the fluctuations
about a mean-field magnetization in the $\alpha=1$ direction.
Inserting (\ref{dual1}) in (\ref{cal}) we find that the action
changes to
\begin{equation}
{\cal L} = \frac{1}{2} \sum_{\alpha=1}^{n} \left[
\left( \nabla_x \widetilde{\Phi}_{\alpha} \right)^2 + \delta_{\alpha,1}  2 |\widetilde{R}|
\widetilde{\Phi}_{\alpha}^2 \right] + \sqrt{\frac{|\widetilde{R}| U}{6}}
\widetilde{\Phi}_{1} \sum_{\alpha=1}^{n} \widetilde{\Phi}_{\alpha}^2 +
\frac{U}{24} \left( \sum_{\alpha=1}^{n} \widetilde{\Phi}_{\alpha}^2 \right)^2
\label{dual1a}
\end{equation}
This action will form the basis for a perturbation theory in $U$
(or equivalently by (\ref{s3}), in ${\cal G}_D$)
in the rest of this appendix.
Notice that there is a cubic term in this action, and so $\langle \widetilde{\Phi}_{1}
\rangle \neq 0$. Indeed, the expression for the incipient `spontaneous
magnetization' $N_0 \equiv \langle \Phi_1 \rangle$, correct to
zeroth order in $U$, can be obtain in a straightforward
perturbative calculation from (\ref{dual1a}), and we obtain:
\begin{equation}
N_0 =\sqrt{ \frac{6 |\widetilde{R}|}{U}} \left[
1 - \frac{T U}{4 |\widetilde{R}|} \int^{1/a} \frac{d^2 k}{(2
\pi)^2} \left( \frac{1}{k^2 + 2 |\widetilde{R}|} + \frac{n-1}{3
k^2} \right) \right].
\label{dual2}
\end{equation}
The fluctuation corrections above come from directions
longitudinal and transverse to the incipient magnetization respectively. Notice
that the latter lead to an infrared divergence for $n>1$, which
correctly suggest that the transverse fluctuations are so strong
that the actual spontaneous magnetization is 0. For now we will
ignore this infrared divergence, as we will see that this
divergence disappears upon insertion of (\ref{dual2}) into
combinations which are ${\rm O}(n)$ invariant.
To complete the computation of the ${\rm O}(n)$-invariant two point $\Phi_{\alpha}$
correlator, we need the correlations of $\widetilde{\Phi}_{\alpha}$. It is
sufficient to compute these at tree level, and we have the formal
expression
\begin{eqnarray}
\left \langle \widetilde{\Phi}_{\alpha} (x) \widetilde{\Phi}_{\alpha} (0) \right
\rangle = T \int^{1/a} \frac{d^2 k}{ (2 \pi )^2} \frac{e^{i k
x}}{k^2 + 2 |\widetilde{R}| \delta_{\alpha,1}},
\label{dual3}
\end{eqnarray}
where there is no implied sum over $\alpha$ on the left hand side.
We can now combine (\ref{dual1}), (\ref{dual2}), and (\ref{dual3})
to obtain an expression for $\sum_{\alpha=1}^{n}
\langle \Phi_{\alpha} (x) \Phi_{\alpha} (0) \rangle$ correct to zeroth order in $U$.
In this expression
we express $\widetilde{R}$ in terms of $R_D$ using (\ref{defQ}), and
also expand the resulting expression consistently to zeroth order
in $U$. Finally, using the definition of ${\cal G}_D$ in (\ref{s3}), we obtain
one of the main results of this Appendix:
\begin{equation}
\sum_{\alpha=1}^{n}
\langle \Phi_{\alpha} (x) \Phi_{\alpha} (0) \rangle =
\frac{3 T}{{\cal G}_D} \left[1 + \frac{{\cal G}_D}{3}
\int \frac{d^2 k}{(2 \pi)^2} \left(
(n-1) \frac{e^{ikx} - 1}{k^2} + \frac{e^{ikx} + n - 1}{k^2 + R_D}
\right) + {\cal O} ({\cal G}_D^2) \right]
\label{dual4}
\end{equation}
It is simple to check the crucial property that (\ref{dual4}) is free of {\em both}
ultraviolet and infrared divergences; the former arises because
the renormalization in (\ref{defQ}) is expected to cure all short
distance problems, and the latter because the left hand side is
$O(n)$-invariant.

The remaining analysis is quite different for different values of
$n$, and we will consider the cases $n=1$, $n=2$ and $n \geq 3$ in the
following subsections.

\subsection{$n=1$}
\label{dualising}

The expression (\ref{dual2}) for the spontaneous magnetization is
free of infra-red divergences for this case; it can also be made
ultra-violet finite by
expressing $\widetilde{R}$ in terms of $R_D$ using (\ref{defQ}).
This procedure leads to the simple result that
\begin{equation}
N_0 = \sqrt{\frac{3T}{{\cal G}_D}} \left[ 1 + {\cal O} ({\cal
G}_D^2) \right].
\label{dual4a}
\end{equation}
The expression (\ref{dual4}) shows that the two-point
$\Phi_{\alpha}$ correlator approaches $N_0^2$ exponentially over a
length scale $1/\sqrt{R_D}$.

It is not difficult to extend (\ref{dual4a}) to obtain the
two-loop ${\cal O} ({\cal G}_D^2 )$ correction by evaluating
$\langle \widetilde{\Phi}_1 \rangle$ in a perturbation theory in $U$ under
the action (\ref{dual1a}). We verified that by
expressing $\widetilde{R}$ in terms of $R_D$ using (\ref{defQ}),
and expanding consistently to the needed order in $U$,
all the ultraviolet divergent terms in the resulting expression
cancelled,
and we obtained
\begin{eqnarray}
N_0 = \sqrt{\frac{3T}{{\cal G}_D}} && \left[
1 - \frac{T^2 U^2}{12 R_D}
\int \frac{d^2 k}{(2 \pi)^2} \int \frac{d^2 k}{(2 \pi)^2}
\left\{ \frac{9 R_D}{(k^2 + R_D)^2 (p^2 + R_D)
((\vec{k}+\vec{p})^2 + R_D)} \right. \right. \nonumber \\
&&~~~~~~~~~~~~~~~~+ \left. \left. \frac{2}{(k^2 + R_D) (p^2 + R_D)
((\vec{k}+\vec{p})^2 + R_D)} \right\}  + {\cal O} (U^3) \right].
\label{dual4b}
\end{eqnarray}
The integrals in (\ref{dual4b}) are easily evaluated to yield
\begin{equation}
N_0 = \sqrt{\frac{3T}{{\cal G}_D}} \left[
1 - \frac{5 {\cal G}_D^2}{12} J(1,1,1) + {\cal O} ( {\cal G}_D^3 )
\right],
\label{dual4c}
\end{equation}
where the numerical constant $J(1,1,1)$ is given later in
(\ref{dual5i}).

\subsection{$n=2$}
\label{dualxy}

Evaluating (\ref{dual4}) for $x \gg 1/\sqrt{R_D}$ and $n=2$ we obtain
\begin{equation}
\sum_{\alpha=1}^n \langle \Phi_{\alpha} (x) \Phi_{\alpha} (0) \rangle =
\frac{3 T}{{\cal G}_D} \left[1 - \frac{{\cal G}_D}{6 \pi}
\left( \ln (|x|\sqrt{R_D}) - \ln2 + \gamma \right) + \ldots
\right],
\label{dual5}
\end{equation}
where $\gamma$ is Euler's constant.
We are now in the magnetically ordered state of the $n=2$ XY
model, and we know this system has power-law decay of spatial
correlations. Therefore we assert that (\ref{dual5})
exponentiates into the result (\ref{s8}), and identify the spin
stiffness $\rho_s (T)$ as
\begin{equation}
\frac{\rho_s (T)}{T} = \frac{3}{{\cal G}_D} + {\cal O} ({\cal G}_D^0).
\label{dual5a}
\end{equation}

We will now compute the next two terms in the small ${\cal G}_D$
expansion for $\rho_s (T)$ in (\ref{dual5a}). Rather than
obtaining these by computing higher order corrections to the
correlator in (\ref{dual5}), we will compute the stiffness
directly by examining the response of the free energy density to
an external field $\vec{H}$ which couples to the generator of $O(2)$
rotations. This is equivalent to computing the change in the free
energy in the presence of twisted boundary conditions. In
particular, we modify the action in (\ref{cal}) by the
substitution
\begin{equation}
(\nabla_x \Phi_{\alpha})^2 \Longrightarrow (\nabla_x \Phi_1 - i
\vec{H} \Phi_2)^2 + (\nabla_x \Phi_2 + i
\vec{H} \Phi_1)^2 ,
\label{dual5b}
\end{equation}
and compute the small $H$ dependence of the free energy $\ln {\cal Z}
(H)$. The stiffness is defined by
\begin{equation}
\frac{1}{V} \ln \frac{{\cal Z}(H)}{{\cal Z} (0)} =  \frac{\rho_s
(T)}{2 T} H^2 + ~~\cdots
\label{dual5c}
\end{equation}
where $V$ is the volume of the system, and
the ellipses represent terms higher order in $H$. We compute
this expression by first modifying (\ref{dual1}) to account for
the $H$ dependence of the mean-field magnetization:
\begin{equation}
\Phi_{\alpha} (x)  = \sqrt{\frac{6 (|\widetilde{R}|+H^2)}{U}} \delta_{\alpha,1}
+ \widetilde{\Phi}_{\alpha} (x),
\label{dual5d}
\end{equation}
and expanding the resulting action in powers of $H$. In this
manner it is not hard to show that to order $H^2$
\begin{eqnarray}
&& \frac{1}{V H^2} \ln \frac{{\cal Z}(H)}{{\cal Z} (0)} =
\frac{3 |\widetilde{R}|}{T U} - \left\langle \widetilde{\Phi}_1^2
(x)
\right \rangle - \sqrt{\frac{U}{24 |\widetilde{R}|}} \left\langle
\widetilde{\Phi}_1 (x) \left( \widetilde{\Phi}_1^2 (x) + \widetilde{\Phi}_2^2 (x) \right)
\right\rangle \nonumber \\
&&- \frac{1}{4} \int d^2 y \left\langle
\left( \widetilde{\Phi}_1 (y) \vec{\nabla}_y \widetilde{\Phi}_2 (y)
- \widetilde{\Phi}_2 (y) \vec{\nabla}_y \widetilde{\Phi}_1 (y)
\right) \cdot \left( \widetilde{\Phi}_1 (x) \vec{\nabla}_x \widetilde{\Phi}_2 (x)
- \widetilde{\Phi}_2 (x) \vec{\nabla}_x \widetilde{\Phi}_1 (x)
\right) \right\rangle,
\label{dual5e}
\end{eqnarray}
where all expectation values are to be evaluated under the action
(\ref{dual1a}). It is a straightforward, but lengthy, exercise to compute the right hand
side of (\ref{dual5e}) in a perturbation theory in $U$, which
requires enumerating all Feynman graphs to two loops.
After formal expressions for the graphs have been obtained, we
perform the substitution (\ref{defQ}) to replace $\widetilde{R}$ by $R_D$,
and again collect terms to
order $U^1$.
We will not
present the details of this here, but will state the
result obtained without any further manipulations on the
expressions for the individual Feynman graphs:
\begin{eqnarray}
&& \frac{\rho_s (T)}{2T} = \frac{3R_D}{2 T U} + T U \int \frac{d^d p}{(2
\pi)^d} \int \frac{d^d k}{ (2 \pi)^d} \Biggl\{ \nonumber \\
&&~~~ -\frac{R_D}{3(p^2 + R_D)^2} \left[
\frac{2}{k^2 (k^2 + R_D)} + \frac{9}{(k^2
+ R_D)( (\vec{k} + \vec{p})^2 + R_D)} + \frac{1}{k^2
(\vec{k} + \vec{p})^2} \right] \nonumber \\
&&~~~ -\frac{R_D}{3p^2 (p^2 + R_D)} \left[
- \frac{1}{k^2 (k^2 + R_D)} + \frac{1}{k^2 ( (\vec{k} + \vec{p})^2 + R_D)}  \right] \nonumber \\
&&~~~ + \frac{1}{6 ( p^2 + R_D)} \left[
\frac{1}{k^2 (\vec{p}+\vec{k})^2} + \frac{3}{(k^2+R_D) ((\vec{p}+\vec{k})^2+R_D)}
\right] \nonumber \\
&&~~~+ \left. \frac{1}{3 p^2 k^2 (p^2 + R_D)(k^2 + R_D)} \left[
\frac{3 \vec{p} \cdot \vec{k}}{((\vec{p}+ \vec{k})^2 + R_D)}
- \frac{\vec{p} \cdot \vec{k}}{(\vec{p}+\vec{k})^2} \right]
\right \}.
\label{dual5f}
\end{eqnarray}
It is directly apparent that all terms in (\ref{dual5f}) are
individually ultraviolet convergent: this has been achieved, as
expected, by the substitution of $\widetilde{R}$ by $R_D$. However,
many of the terms infrared divergent, and it is
not at all apparent that such divergences will cancel between the
various terms. To control these divergences, we evaluate the
terms in dimensional regularization {\em i.e.} the $p$ and $k$
integrals are evaluated in a dimension $d$ just above 2, and the
resulting terms expanded in a series in $(d-2)$. We show below
that while there are individual terms of order $1/(d-2)^2$ and
$1/(d-2)$, they do cancel among each other.

By a series of elementary algebraic manipulations (including splitting apart
some of the denominators by the method of partial fractions), all the terms
in (\ref{dual5f}) can be expressed in terms of two basic integral
expressions. These are
\begin{equation}
I(a,b) \equiv \int \frac{d^d k}{(2 \pi)^d}
 \frac{1}{(k^2 + a) (k^2 + b)},
\end{equation}
and
\begin{equation}
J(a,b,c) \equiv \int \frac{d^d k}{(2 \pi)^d} \int \frac{d^d p}{(2
\pi)^d} \frac{1}{(k^2 + a) (p^2 + b) ((\vec{k}+\vec{p})^2 + c)};
\label{dual5g}
\end{equation}
it is easy to see that both $I$ and $J$ are invariant under all
permutations of their arguments {\em i.e.} $I(a,b)=I(b,a)$, $J(a,b,c)=J(c,a,b)=J(c,b,a)$
etc. In terms of $I$ and $J$, the result (\ref{dual5f}) takes the
form
\begin{eqnarray}
 \frac{\rho_s (T)}{2T} = && \frac{3R_D}{2 T U} + T U \left\{
\frac{2R_D}{3} I(R_D,0) I(R_D,0) - \frac{2R_D}{3} I(R_D,R_D) I(R_D,0) -
\frac{1}{3} J(R_D,0,0)  \right. \nonumber \\
&&\left.+ J(R_D,R_D,R_D) + R_D \frac{d}{d R_D} J(R_D,R_D,R_D)
+ \frac{R_D}{3} \frac{d}{d R_D}
J(R_D,0,0) \right\}
\label{dual5h}
\end{eqnarray}
We can evaluate the needed values of $I$ and $J$ by standard
methods which transform the momentum integrals into integrals over
Feynman parameters~\cite{ramond}:
\begin{eqnarray}
I(R_D,R_D) &=& \frac{\Gamma(2-d/2)}{(4 \pi)^{d/2} R_D^{2-d/2}} \nonumber
\\
I(R_D,0) &=& \frac{2 I(R_D,R_D)}{(d-2)}  \nonumber
\\
J(R_D,0,0) &=& \frac{2 \Gamma(3-d) \Gamma(d/2 - 1) \Gamma(2-d/2)}{(4
\pi)^{d} R_D^{3-d} (d-2)} \nonumber \\
J(R_D,R_D,R_D) &=& \frac{1}{16 \pi^2 R_D} \int_0^1 dy_1 \int_0^1 dy_2
\frac{1}{y_2 + y_1 (1-y_1)(1-y_2)} \nonumber \\
&=& \frac{2.34390723869}{16 \pi^2 R_D}.
\label{dual5i}
\end{eqnarray}
The last two expressions have been given directly in $d=2$, as $J(R_D,R_D,R_D)$ does
not have any poles in $(d-2)$ and also does not appear in
combinations multiplying poles in (\ref{dual5h}). Inserting
(\ref{dual5i}) into (\ref{dual5h}) and expanding in powers of
$(d-2)$, we find that all double poles and poles in $(d-2)$ cancel, and we obtain
our final result generalizing (\ref{dual5a}):
\begin{equation}
\frac{\rho_s (T)}{T} = \frac{3}{{\cal G}_D} - \frac{{\cal G}_D}{36}
+ {\cal O} ({\cal G}_D^2).
\label{dual5j}
\end{equation}

\subsection{$n \geq 3$}
\label{dualheis}
The argument is more subtle for these cases with non-Abelian
symmetry. We now expect that at length scales $\gg 1/\sqrt{R_D}$
all longitudinal fluctuations will freeze out, and the transverse
fluctuations will map onto those ${\rm O} (n)$ non-linear sigma
model. This model has a dimensionless coupling constant $g$, and
for $g\ll 1$ there is a large correlation length of order (see {\em
e.g.} Ref.~\onlinecite{CHN,CSY})
\begin{equation}
\xi \sim  \frac{1}{\sqrt{R_D}}
 \left[\frac{(n-2)g}{2 \pi} \right]^{1/(n-2)} \exp\left( \frac{2
\pi}{(n-2)g} \right),
\label{dual6}
\end{equation}
where we have chosen $1/\sqrt{R_D}$ as the natural short distance
cutoff of the non-linear sigma model.
For $1/\sqrt{R_D} \ll |x| \ll \xi$, the two
point-correlations behave as~\cite{CSY}
\begin{eqnarray}
\left \langle \Phi_{\alpha} (x) \Phi_{\alpha} ( 0) \right \rangle
&\propto & \left[(n-2)g \ln \left( \frac{\xi}
{|x|} \right) \right]^{(n-1)/(n-2)} \nonumber \\
&\propto & \left[ 1 - \frac{(n-1)g}{2 \pi} \ln \left(|x|\sqrt{R_D}\right) + \ldots \right].
\label{dual7}
\end{eqnarray}
Let us compare this with the expression obtained from
(\ref{dual4}), which yields in the same regime
\begin{equation}
\left \langle \Phi_{\alpha} (x) \Phi_{\alpha} ( 0) \right \rangle
= \frac{3 T}{{\cal G}_D} \left[1 - \frac{(n-1){\cal G}_D}{6 \pi} \ln \left(
|x| \sqrt{R_D} \right) + \ldots \right].
\label{dual8}
\end{equation}
Comparing (\ref{dual7}) and (\ref{dual8}), we can obtain the missing
prefactor in (\ref{dual7}), and also get
\begin{equation}
g = \frac{{\cal G}_D}{3}
\label{dual9}
\end{equation}

Actually it is possible to do better, and actually fix the missing
constant in (\ref{dual6}) precisely.
To do this, as was shown by Hasenfratz and Niedermayer~\cite{HN},
we need to carry out exactly the calculation of
Section~\ref{dualxy} and obtain the $H$ dependence of the free
energy density for $n \geq 3$.
Now the analog of the replacement (\ref{dual5b}) in the action
(\ref{cal}) is
\begin{equation}
(\nabla_x \Phi_{\alpha})^2 \Longrightarrow (\nabla_x \Phi_1 - i
\vec{H} \Phi_2)^2 + (\nabla_x \Phi_2 + i
\vec{H} \Phi_1)^2 + \sum_{\alpha > 2} (\nabla_x \Phi_{\alpha})^2 .
\label{dual10}
\end{equation}
Next we will compute the $H$ dependence of ${\cal Z}(H)$ in a
perturbation theory in $U$, but will find that the structure of
the answer is actually quite different from that found in
(\ref{dual5f}) for $n=2$. In the present situation one finds that
all the infrared divergences in $(1/H^2) \ln {\cal Z} (H)/{\cal Z}(0)$
do {\em not} cancel, which correctly indicates that the
renormalized stiffness $\rho_s (T)$ is strictly zero for all
$T>0$. Instead, the small $H$ dependence of ${\cal Z}(H)$ is more
complex, and we will show that $\ln {\cal Z}(H)/{\cal Z}(0)
\sim H^2 \ln(1/H)$.

After substitution of (\ref{dual10}) into (\ref{cal}), it is
immediately apparent that at order $U^0$ ${\cal Z}(H)$ consists of
two separate contributions. The first is the contribution of the $\widetilde{\Phi}_{1,2}$
components and this is identical to that computed in
Section~\ref{dualxy} for $n=2$, The second is the contribution of
the remaining $n-2$ components which, at this order, are simply
free fields with `mass' $H^2$. So we get
\begin{equation}
\ln \frac{{\cal Z}(H)}{{\cal Z}(0)} = H^2 \left[
\frac{3|\widetilde{R}|}{T U}
+ \int \frac{d^2 p}{(2 \pi)^2} \frac{2}{p^2 + 2 |\widetilde{R}|}
\right] - \frac{(n-2)}{2} \int \frac{d^2 p}{(2 \pi)^2} \ln \left( \frac{p^2
+H^2}{p^2} \right),
\label{dual10a}
\end{equation}
where the first two terms are obtained by evaluating
(\ref{dual5e}) to order $U^0$, and the last term is the
contribution of the remaining $(n-2)$ components. Now substituting
$R_D$ for $\widetilde{R}$ by using (\ref{defQ}) we obtain, as
expected, an expression free of ultraviolet divergences:
\begin{equation}
\ln \frac{{\cal Z}(H)}{{\cal Z}(0)} =
\frac{3R_D H^2 }{2T U} -  \frac{(n-2)}{2}
\int \frac{d^2 p}{(2 \pi)^2} \left[ \ln \left( \frac{p^2
+H^2}{p^2} \right) - \frac{H^2}{p^2 + R_D} \right],
\label{dual10b}
\end{equation}
Finally, we evaluate the integral in the limit $H^2 \ll R_D$, and
express the result in terms of ${\cal G}_D$ using (\ref{s3}):
\begin{equation}
\ln \frac{{\cal Z} (H)}{{\cal Z} (0)} =  H^2 \left[ \frac{3}{2
{\cal G}_D} + \frac{(n-2)}{4 \pi} \ln \left( \frac{H}{(e R_D)^{1/2}} \right) + \ldots \right]
\label{dual11}
\end{equation}
We can now deduce the correlation length, $\xi$, from this
result using the matching to the Bethe ansatz solution, as
discussed in Ref~\onlinecite{HN}: the result is
\begin{equation}
\xi =  \frac{1}{\sqrt{R_D}} \Gamma\left(\frac{n-1}{n-2} \right)
  \left[\frac{e(n-2)g}{16 \pi} \right]^{1/(n-2)} \exp\left(
 \frac{2
\pi}{(n-2)g} \right)
\label{dual12}
\end{equation}
where $\Gamma(x)$ is the gamma function and
$g$ is given in (\ref{dual9}).




\begin{references}

\bibitem{aeppliscience} G.~Aeppli, T.~E.~Mason, S.~M.~Hayden,
H.~A.~Mook, and J.~Kulda, Science {\bf 278}, 1432 (1998).

\bibitem{tran} J.~M.~Tranquada, J.~D.~Axe, N.~Ichikawa, A.~R.~Moodenbaugh,
Y.~Nakamura and S.~Uchida, Phys. Rev. Lett. {\bf 78}, 338 (1997).

\bibitem{DSZ} S.~Das Sarma, S.~Sachdev, and L.~Zheng,
Phys. Rev. B {\bf 58}, 4672 (1998).

\bibitem{pelle} V.~Pellegrini, A.~Pinczuk, B.~S.~Dennis, A.~S.~Plaut,
L.~N.~Pfeiffer, and K.~W.~West,
Science {\bf 281}, 799 (1998).

\bibitem{hardy} W.~N.~Hardy, S.~Kamal, R.~Liang, D.~A.~Bonn, C.~C.~Homes, D.~N.~Basov,
and T.~Timusk in {\it Proceedings
of the 10th Anniversary HTS Workshop
on Physics, Materials and Applications\/} edited by B.~Batlogg {\em et al.}
(World Scientific, Singapore), p. 223 (1996).

\bibitem{eps} S.~Sachdev, Phys. Rev. B {\bf 55}, 142 (1997).

\bibitem{nandini} N.~Trivedi and M.~Randeria, Phys. Rev. Lett.
{\bf 75}, 312 (1995).

\bibitem{ek} V.~J.~Emery and S.~A.~Kivelson, Nature {\bf 374}, 434
(1995); V.~J.~Emery and S.~A.~Kivelson, J. Phys. Chem Solids
{\bf 59}, 1705 (1998).

\bibitem{franz} M.~Franz and A.~J.~Millis, Phys. Rev. B {\bf 58},
14572 (1998).

\bibitem{dorsey} H.-J.~Kwon and A.~T.~Dorsey, cond-mat/9809225.

\bibitem{ds} K.~Damle and S.~Sachdev, Phys. Rev. B {\bf 56}, 8714 (1997).

\bibitem{SY} S.~Sachdev, and J.~Ye  Phys. Rev. Lett. {\bf 69}, 2411 (1992).

\bibitem{CSY} A.~V.~Chubukov, S.~Sachdev, and J.~Ye
Phys.  Rev. B {\bf 49}, 11919 (1994).

\bibitem{ramond} P.~Ramond, {\it Field Theory, A Modern Primer}
(Benjamin-Cummings, Reading, 1981).

\bibitem{loinaz} W.~Loinaz and R.~S.~Willey, Phys. Rev. D {\bf
58}, 076003 (1998).

\bibitem{hhm} B.~I.~Halperin, P.~C.~Hohenberg, and S.~k.~Ma,
Phys. Rev. Lett. {\bf 29}, 1548 (1972);
Phys. Rev. B {\bf 10}, 139 (1974).

\bibitem{halphoh} P.~C.~Hohenberg and B.~I.~Halperin, Rev. Mod. Phys.
{\bf 49}, 435 (1977).

\bibitem{chang} S.-J.~Chang, Phys. Rev. D {\bf 13}, 2778 (1976).

\bibitem{CHN} S.~Chakravarty, B.~I.~Halperin, and D.~R.~Nelson,
Phys. Rev. B {\bf 39}, 2344 (1989).

\bibitem{tyc} S.~Tyc, B.I.~Halperin and S.~Chakravarty, Phys. Rev. Lett
{\bf 62}, 835 (1989).

\bibitem{toral} R.~Toral and A.~Chakrabarti, Phys. Rev. B {\bf
42}, 2445 (1990).

\bibitem{wolff} U.~Wolff, Phys. Rev. Lett. {\bf 62}, 361 (1989).

\bibitem{QXY} K. Harada and N. Kawashima, Phys. Rev. B {\bf 55},
R11949 (1997); J. Phys. Soc. Jpn. {\bf 67}, 2768 (1998).

\bibitem{Weber} H. Weber and P. Minnhagen, Phys. Rev. B {\bf 37},
5986 (1987).

\bibitem{leggett} A.~J.~Leggett, S.~Chakravarty, A.~T.~Dorsey, M.~P.~A.~Fisher,
A.~Garg, and W.~Zwerger, {\it Rev. Mod. Phys.} {\bf 59}, 1 (1987).

\bibitem{dissqm} U.~Weiss, {\em Quantum Dissipative Systems}
(World Scientific, Singapore, 1993).

\bibitem{lesage} F.~Lesage and H.~Saleur, {\it Nucl. Phys. B}
{\bf 493}, 613 (1997).

\bibitem{seoul} S.~Sachdev in {\it Highlights in Condensed Matter
Physics}, Y.~M.~Cho and M.~Virasoro eds. (World Scientific,
Singapore); cond-mat/9811110.

\bibitem{RS} N.~Read and S.~Sachdev, Phys. Rev. Lett. {\bf 62}, 1694 (1989).

\bibitem{keimer} B.~Keimer, H.~F.~Fong, S.~H.~Lee, D.~L.~Milius
and I.~A.~Aksay, cond-mat/9705103.

\bibitem{zhang} S.~Zhang, Science {\bf 275}, 1089 (1997).

\bibitem{normand} B.~Normand and T.~M.~Rice, {\it Phys. Rev. B} {\bf 54}, 7180
(1996); {\em ibid} {\bf 56}, 8760 (1997).

\bibitem{HN} P.~Hasenfratz, M.~Maggiore, and F.~Niedermayer,
Phys. Lett. B {\bf 245}, 522 (1990); P.~Hasenfratz and F.~Niedermayer,
Phys. Lett. B {\bf 245}, 529 (1990); P.~Hasenfratz and F.~Niedermayer,
Phys. Lett. B {\bf 268}, 231 (1991).

\end{references}
\end{document}